%% file: MOJAVE_paperX.tex
\documentclass[iop]{emulateapj}
\usepackage{apjfonts} 

\usepackage{graphicx}
\usepackage{verbatim}

\newcommand{\n}{\nodata}
\newcommand{\be}{\begin{itemize}}
\newcommand{\ee}{\end{itemize}}
\newcommand{\muasyr}{\hbox{$\; \mu{\rm as \ y}^{-1}\;$}}

\def\fermi{\textit{Fermi }}
\def\gr{$\gamma$-ray }

\makeindex
\citeindextrue

\received{June 6, 2013}
\accepted{August 5, 2013}
\journalinfo{Astronomical~journal}
\submitted{}

\shorttitle{MOJAVE. X.}
\shortauthors{M. L. Lister et al.}

\begin{document}

\title{MOJAVE. X. Parsec-Scale Jet Orientation Variations and Superluminal Motion in AGN}

\author{ M. L. Lister\altaffilmark{1},
M. F. Aller\altaffilmark{2},
H. D. Aller\altaffilmark{2},
D. C. Homan\altaffilmark{3},
K. I. Kellermann\altaffilmark{4},
Y. Y. Kovalev\altaffilmark{5,6},
A. B. Pushkarev\altaffilmark{7,8,6},
J. L. Richards\altaffilmark{1},
E. Ros\altaffilmark{9,10,6},
T. Savolainen\altaffilmark{6}
}

\altaffiltext{1}{
Department of Physics, Purdue University, 525 Northwestern Avenue,
West Lafayette, IN 47907, USA;
\email{mlister@purdue.edu}
}
\altaffiltext{2}{
Department of Astronomy, University of Michigan, 817 Dennison Building, Ann Arbor, MI 48 109, USA;
}

\altaffiltext{3}{
Department of Physics, Denison University, Granville, OH 43023;}

\altaffiltext{4}{
National Radio Astronomy Observatory, 520 Edgemont Road, Charlottesville, VA 22903, USA;
}

\altaffiltext{5}{
Astro Space Center of Lebedev Physical Institute,
Profsoyuznaya 84/32, 117997 Moscow, Russia;
}
\altaffiltext{6}{
Max-Planck-Institut f\"ur Radioastronomie, Auf dem H\"ugel 69,
53121 Bonn, Germany;
}
\altaffiltext{7}{
Pulkovo Observatory, Pulkovskoe Chaussee 65/1, 196140 St.
Petersburg, Russia;
}
\altaffiltext{8}{
Crimean Astrophysical Observatory, 98409 Nauchny, Crimea, Ukraine;}


\altaffiltext{9}{
Observatori Astron\`omic, Universitat de Val\`encia,
  Parc Cient\'{\i}fic, C. Catedr\'atico Jos\'e Beltr\'an 2, E-46980
  Paterna, Val\`encia, Spain}

\altaffiltext{10}{
Departament d'Astronomia i Astrof\'{\i}sica,
  Universitat de Val\`encia, C. Dr. Moliner 50, E-46100 Burjassot,
  Val\`encia, Spain}

\begin{abstract}
  We describe the parsec-scale kinematics of 200 AGN jets based on 15
  GHz VLBA data obtained between 1994 Aug 31 and 2011 May 1. We 
  present new VLBA 15 GHz images of these and 59 additional AGN from
  the MOJAVE and 2 cm Survey programs. Nearly all of the 60 most
  heavily observed jets show significant changes in their innermost
  position angle over a 12 to 16 year interval, ranging from
  $10\arcdeg$ to $150\arcdeg$ on the sky, corresponding to intrinsic
  variations of $\sim 0.5\arcdeg$ to $\sim 2\arcdeg$. The BL Lac jets
  show smaller variations than quasars.  Roughly half of the heavily
  observed jets show systematic position angle trends with time, and
  20 show indications of oscillatory behavior.  The time spans of the
  data sets are too short compared to the fitted periods (5 to 12 y),
  however, to reliably establish periodicity. The rapid changes and
  large jumps in position angle seen in many cases suggest that the
  superluminal AGN jet features occupy only a portion of the entire
  jet cross section, and may be energized portions of thin instability
  structures within the jet.  We have derived vector proper
  motions for 887 moving features in 200 jets having at least five
  VLBA epochs. For 557 well-sampled features, there are sufficient
  data to additionally study possible accelerations.  We find that the
  moving features are generally non-ballistic, with 70\% of the
  well-sampled features showing either significant accelerations or
  non-radial motions.  Inward motions are rare (2\% of all features),
  are slow ($<$ 0.1 mas per y), are more prevalent in BL Lac jets, and
  are typically found within 1 mas of the unresolved core feature.
  There is a general trend of increasing apparent speed with distance
  down the jet for both radio galaxies and BL Lac objects.  In most
  jets, the speeds of the features cluster around a characteristic
  value, yet there is a considerable dispersion in the distribution.
  Orientation variations within the jet cannot fully account for the
  dispersion, implying that the features have a range of Lorentz factor
  and/or pattern speed.  Very slow pattern speed features are rare,
  comprising only 4\% of the sample, and are more prevalent in radio
  galaxy and BL Lac jets.  We confirm a previously reported upper
  envelope to the distribution of speed versus beamed luminosity for
  moving jet features.  Below $10^{26} \; \mathrm{W\;Hz^{-1}}$ there
  is a fall-off in maximum speed with decreasing 15 GHz radio
  luminosity.  The general shape of the envelope implies that the most
  intrinsically powerful AGN jets have a wide range of Lorentz factors
  up to $\sim 40$, while intrinsically weak jets are only mildly
  relativistic.
\end{abstract}
 

\section{INTRODUCTION} 

High resolution multi-epoch radio observations of jetted outflows
associated with active galactic nuclei (AGN) have contributed
substantially to our understanding of the immediate environments of
supermassive black holes, by providing direct measurements of jet flow
kinematics and magnetic field properties.  Since its construction in
1994, the Very Long Baseline Array (VLBA) has been used to regularly
image the brightest radio-loud AGN jets and study their evolution on
parsec-scales \citep{2004ApJ...609..539K}. The VLBA 2cm Survey
\citep{2cmPaperI} sampled the jet kinematics of 110 AGN, and was
succeeded in 2002 by the MOJAVE program, which added full polarization
imaging and defined a complete northern-sky radio flux density-limited
sample \citep[hereafter Paper~I]{MOJAVE_I}.  Kinematic results for 127
MOJAVE jets based on data spanning 1994--2007 were presented by
\citealt[(hereafter Paper~VI)]{MOJAVE_VI} and \citealt[(hereafter Paper
VII)]{MOJAVE_VII}. They showed that bright jet features typically
exhibit apparent superluminal speeds and accelerated motions. These
findings are consistent with the widely accepted picture of high bulk
Lorentz factor jets viewed at angles very close to the line of sight,
i.e., blazars.  Although blazars are quite rare in the general AGN
parent population, their predominance in the flux density-limited
MOJAVE sample is a direct result of Doppler orientation bias
\citep{1982MNRAS.200.1067O}, since the observed flux densities of
aligned, fast jets are highly Doppler boosted by relativistic
aberration effects.

\input{gentablestub}

The MOJAVE program has confirmed an important trend, first reported by
\cite{1995PNAS...9211385V} in the Caltech-Jodrell AGN survey, in
which jets with the fastest superluminal speeds all tend to have high
Doppler boosted radio luminosities.  To first order, such a trend
might be expected from orientation and Doppler boosting effects, but
an analysis by \cite{2007ApJ...658..232C} and Monte Carlo simulations
presented in Paper~VI indicated that there is a correlation between
intrinsic jet speed and intrinsic (de-beamed) luminosity present in
the population. In the absence of such a correlation, we would expect
to see highly superluminal jets at much lower boosted radio
luminosities.

In order to further investigate these issues, we expanded the MOJAVE
program in 2006 to include regular VLBA imaging of additional
low-luminosity AGN jets. In 2009, we expanded the sample again to
encompass new $\gamma$-ray loud blazar jets discovered by {\it Fermi}
\citep{2011ApJ...742...27L}.

We present new VLBA 15 GHz images of the original 135 source MOJAVE
flux-density limited sample obtained between 2007 September 6 and 2011
May 1.  We also present VLBA images of 124 additional AGN from three
new AGN jet samples, based on 15 GHz VLBA data obtained between 1994
Aug 31 and 2011 May 1 from the NRAO archive and the MOJAVE program
\citep[hereafter Paper~V]{MOJAVE_V}.  These include a complete
radio-selected sample above 1.5 Jy, a complete $\gamma$-ray selected
sample, and a representative low-luminosity AGN jet sample.  We use
these data to present an updated kinematics analysis of the 135 jets
in the original MOJAVE flux-density limit sample, and first ever
kinematics analyses of 65 jets for which we have obtained at least 5
VLBA epochs.

The overall layout of the paper is as follows. In
Sections~\ref{samples} and \ref{obs} we discuss the samples and
observational data, respectively.  In Section~\ref{analysis} we
describe our method of modeling the individual jet features and their
kinematic properties. We discuss overall trends in the data in
Section~\ref{discussion}, and summarize our findings in
Section~\ref{conclusions}. We adopt a cosmology with $\Omega_m =
0.27$, $\Omega_\Lambda = 0.73$ and $H_o = 71 \; \mathrm{km\; s^{-1} \;
  Mpc^{-1}}$. We refer to the radio sources throughout using either
B1950 nomenclature or commonly-used aliases, which we list in
Table~\ref{gentable}.

\section{AGN SAMPLE DEFINITIONS}\label{samples}

\subsection{Radio-selected MOJAVE 1.5 Jy sample}

Unlike blazar surveys in the optical or soft X-ray regimes, the radio
emission from the brightest radio-loud blazars is not substantially
obscured by or blended with emission from the host galaxy. A
VLBA-selected sample thus provides a very ``clean'' blazar sample,
namely, one selected on the basis of (beamed synchrotron) jet
emission.  

In Paper~V we described the original radio-selected MOJAVE sample of
135 AGN, which was based on the 15 GHz VLBA flux density exceeding
1.5 Jy (2 Jy for declination $< 0^\circ$) at any epoch during the
period 1994.0--2004.0. The sky region was limited to declination $\ge
-20^\circ$ and the Galactic plane region $|b| < 10^\circ$.  In order
to encompass a broader range of {\it Fermi}-detected AGN, particularly
those which recently entered an active state, we updated our
radio-selection criteria in 2011 to form the complete MOJAVE 1.5 Jy
sample. The latter now consists of all known non-gravitationally
lensed AGN with J2000 declination $> -30^\circ$ (no Galactic plane
restriction) and VLBA flux density $S_{15\;\mathrm{GHz}} > 1.5$ Jy at
any epoch between 1994.0 and 2010.0.  The new list results in a larger
overlap with the {\it Fermi} AGN catalog \citep{2LAC}, and simplifies
the determination of luminosity functions (e.g., \citealt{MOJAVE_IV}),
which are useful for studies of the extragalactic background light and
blazar parent populations.

The overall properties of the sample are summarized in
Table~\ref{gentable}, where the ``R'' notation in column (7) indicates
1.5 Jy radio sample membership. There are 183 AGN in total (see
Table~\ref{surveytable}), with the sample being heavily dominated by
flat spectrum radio quasars (78\%) and BL Lac objects (16\%). The
optical classifications are 98\% complete, with redshifts available for
96\% of the sample. 

\begin{deluxetable}{lccc} 
\tablecolumns{4} 
\tabletypesize{\scriptsize} 
\tablewidth{0pt}  
\tablecaption{\label{surveytable}Optical Classification Summary of AGN Samples}  
\tablehead{\colhead{} & \colhead {1.5 Jy Radio}  &  
\colhead{1FM $\gamma$-ray} &\colhead{Low-Luminosity} } 
\startdata 
Quasars                & 142 & 72 & 7 \\
BL Lacs                &  29 & 42 & 20 \\
Radio Galaxies         &   8 &  1 & 16 \\
Narrow-line Seyfert 1s &   0 &  1 &  0 \\
Unidentified           &   4 &  0 &  0 \\
\cline{1-4}
Total                  & 183 & 116& 43
\enddata
\tablecomments{Some AGNs belong to two or more of the samples listed above (see Fig.~\ref{venn}). }
\end{deluxetable} 

\subsection{$\gamma$-ray-selected 1FM sample}

The continuous all-sky coverage of the LAT instrument on board the
\fermi satellite has significantly improved our knowledge of the
blazar population at $\gamma$-ray energies above 100 MeV by
identifying nearly 1000 AGN associated with $\gamma$-ray sources
\citep{2LAC}. For the joint LAT team-MOJAVE study of
\cite{2011ApJ...742...27L}, we constructed a $\gamma$-ray sample based
on the initial 11-month First {\it Fermi} AGN catalog \citep{1LAC}.
The specific criteria were: i) average integrated $>0.1$ GeV energy
flux $\ge 3 \times 10^{-11} \mathrm{\;erg\; cm^{-2}\; s^{-2}}$ between
2008 August 4 and 2009 July 5, ii) J2000 declination $ > 30\arcdeg$,
iii) galactic latitude $|b| > 10\arcdeg$, and iv) source not
associated with a gravitational lens.  The sample is complete with
respect to $\gamma$-ray flux, with the exception of two $\gamma$-ray
sources (1FGL J1653.6$-$0158 and 1FGL J2339.7$-$0531) which were
dropped since they had no unambiguous radio counterpart.

The overall properties of the sample are described by
\cite{2011ApJ...742...27L} and summarized in Table~\ref{gentable},
where the ``G'' notation in column (7) indicates \gr sample
membership. There are 116 AGN in total (Table~\ref{surveytable}), 56
of which are in common with the MOJAVE 1.5 Jy sample. Like our
radio-selected sample, it is heavily dominated by blazars, but
contains a larger fraction of BL Lac objects (36\%).  The remainder of
the sample are quasars, with the exception of the nearby radio galaxy
NGC~1275 (3C~84) and the narrow-line Seyfert 1 galaxy PMN J0948+0022.

\subsection{Low-luminosity compact AGN sample}

In 2006 we expanded the MOJAVE program to include regular VLBA
observations of 16 AGN with VLBA 15 GHz luminosities below $<
10^{26}\; \mathrm{W \; Hz^{-1}}$. These were chosen from the VLBA
Calibrator Survey \citep{2002ApJS..141...13B,2003AJ....126.2562F,
  2005AJ....129.1163P,2006AJ....131.1872P,2007AJ....133.1236K,VCS6},
based on the following criteria: i) 8 GHz VLBA flux density greater
than 0.35 Jy, ii) $z \le 0.3$, and iii) J2000 declination $>
-30\arcdeg$. By adding the AGN already in the MOJAVE program which met
these criteria, we obtained a final sample of 43 low-luminosity
compact AGN, as indicated by the ``L'' notation in column (7) of
Table~\ref{gentable}.  Although the latter would not typically be
considered as low-luminosity amongst the general radio-loud AGN
population, we will refer to them as the ``low-luminosity'' sample, in
comparison with the typically high luminosity blazars in our radio and
$\gamma$-ray selected samples.  Due to the lack of redshift 
information for the full VLBA Calibrator Survey, our
low-luminosity sample is not complete. However, it is a useful
representative set for examining the kinematics of weaker jets, and in
this respect complements other small-sample low-luminosity AGN VLBI
monitoring programs (e.g.,
\citealt{2001ApJ...552..508G,2010ApJ...723.1150P}).

\begin{figure}[t]
\centering
\includegraphics[trim=5cm 3cm 12cm 3cm,clip,width=0.4\textwidth]{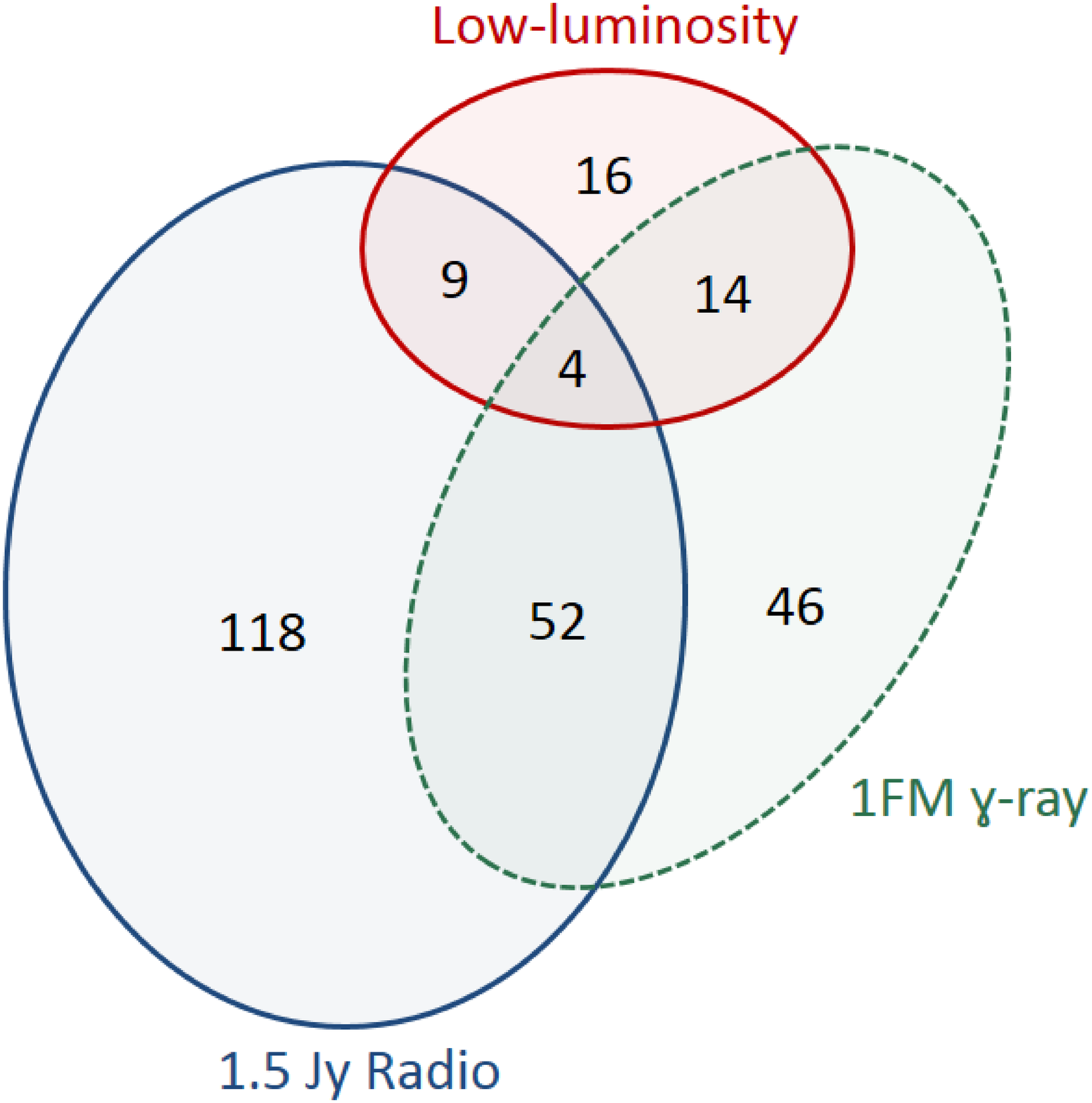}
\caption{\label{venn} Area-proportional Venn diagram with labels
  indicating the total number of AGN in each sub-set of the MOJAVE 1.5 Jy, 1FM
  $\gamma$-ray selected, and low-luminosity AGN samples.  }
\end{figure}

These three samples, comprising 259 AGN in total, provide a broad
cross-section of AGN types among bright, compact radio sources. The
overlap among the samples is shown in Figure~\ref{venn}. Since only
200 of these sources had at least 5 VLBA epochs as of May 2011, the
kinematic data are incomplete. In particular, the AGNs with missing
data tend to be among the weaker radio and $\gamma$-ray selected AGNs
that were added later in the MOJAVE survey. A statistical
inter-comparison of these samples will be presented in future papers
in this series once a full, unbiased dataset has been collected.

\section{VLBA OBSERVATIONS AND DATA REDUCTION}\label{obs}

\input{maptablestub}

\begin{figure*}[p]
\centering
\epsscale{0.92}
\plotone{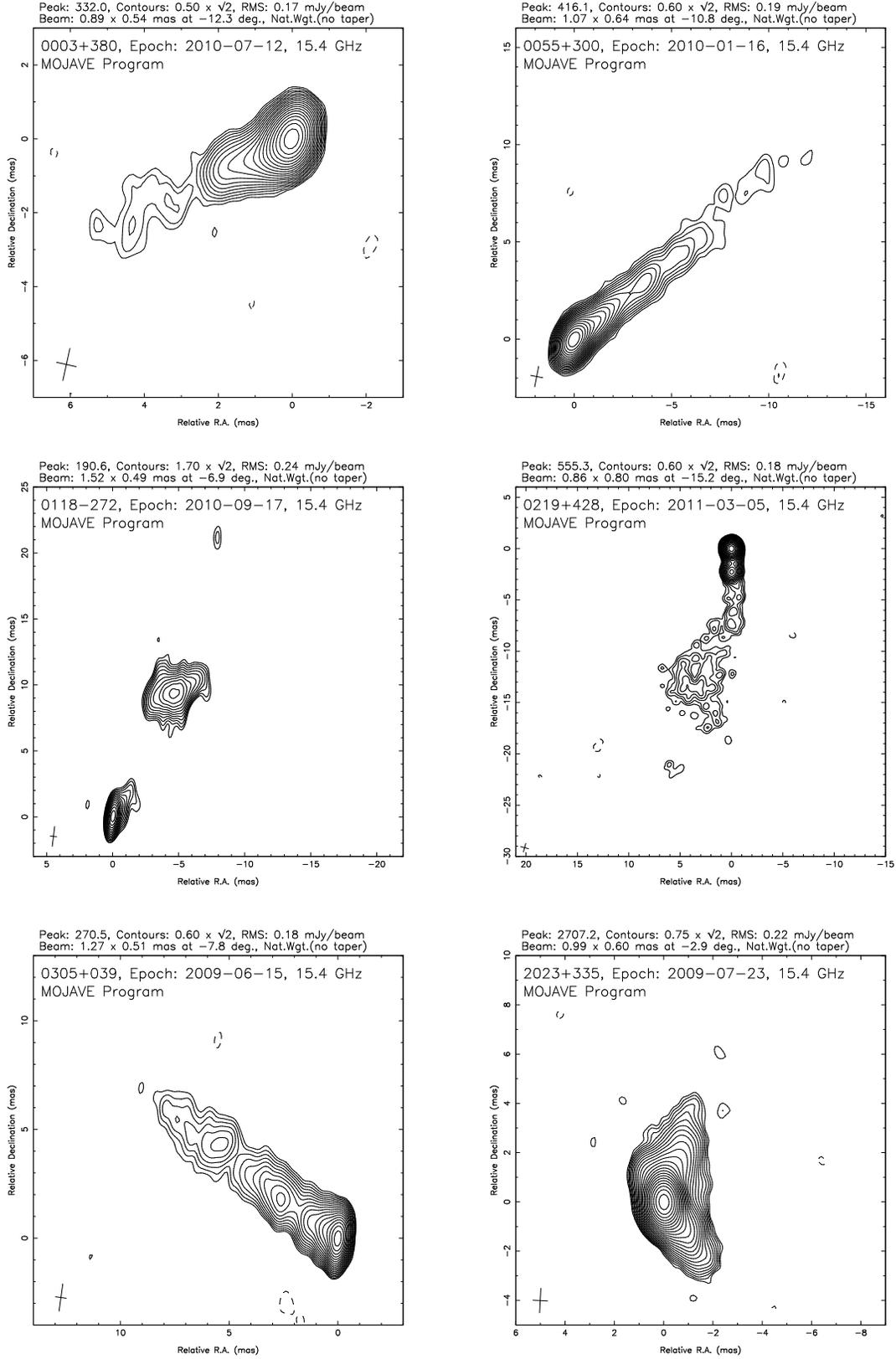}
\caption{\label{images} Naturally-weighted 15 GHz total
  intensity VLBA contour images of individual epoch observations of
  the MOJAVE AGN sample. The contours are in successive powers of
  $\sqrt{2}$ times the base contour level, as listed in
  Table~\ref{maptable} and at the top of each panel. Because of
  self-calibration, in some cases the origin may be coincident with
  the brightest feature in the image, rather than the putative core
  feature listed in Table~\ref{3gaussiantable}.  (This is a figure
  stub, an extended version is available online).  }
\end{figure*}

In Paper~V, we presented 15 GHz VLBA images of the 135 AGN in the
original MOJAVE flux density limited sample based on data from the
MOJAVE programs spanning 1994 Aug 31 to 2007 Sept 9, the VLBA 2m
Survey, and the NRAO archive\footnote{http://archive.nrao.edu/}.  In
Figure~\ref{images} we show naturally-weighted 15 GHz VLBA images
derived from newly acquired VLBA data on these AGN up to 2011 May 1,
as well as VLBA data from 1994 Aug 31 to 2011 May 1 on 124 AGN in our
new samples. 

The multi-epoch observations for each AGN, along with the
corresponding image parameters, are listed in Table~\ref{maptable}.
Column 3 gives the VLBA project code for each observation, along with
an indicator as to whether it is from the MOJAVE program, the VLBA 2
cm Survey, or the NRAO archive. For the latter, we considered only
archival data with at least 4 scans spanning a range of hour angle,
and which included 8 or more VLBA antennas. The VLBA 2 cm Survey
observations (1994--2002) analyzed by \cite{2004ApJ...609..539K}
consist of approximately one hour integrations on each AGN, broken up
into approximately 6--8 minute scans separated in hour angle to
improve the interferometric coverage. A similar observing method and
integration times were used in the full polarization MOJAVE
observations from 2002 May to 2007 September (VLBA codes BL111, BL123,
BL137, and BL149; see Table~\ref{maptable}), and are described in
Paper~I.  During 2006 (VLBA code BL137), the 15 GHz integration times
were shortened by a factor of $\sim 3$ to accommodate interleaved
scans at three other observing frequencies (8.1, 8.4, 12.1 GHz). The
latter data were recently presented by \cite{2012AJ....144..105H} and
\cite{2012AA...545A.113P}.  The MOJAVE and 2 cm Survey observations
were recorded at a data rate of 128 Mbps, which was increased to 256
Mbps in the epochs from 2007 July 3 to 2008 Sept 12 inclusive, and 512
Mbps thereafter.  Beginning with the 2007 Jan 6 epoch, we
increased the number of AGN observed in each 24 hour MOJAVE session
from 18 to 25 AGN to accommodate our expanded monitoring sample
described in Section~\ref{samples}. On 2009 Feb 25 we increased this
further to 30 AGN per session.

\input{gaussiantablestub}

\section{DATA ANALYSIS}\label{analysis}
\subsection{Gaussian Model Fitting}\label{gaussianfitting}

As in Paper~VI, we modelled the $(u,v)$ visibility data at all AGN
epochs using a series of Gaussian components in the Difmap software
package \citep{difmap}. In the majority of cases, we used circular
Gaussians for jet features, and occasionally (when necessary)
elliptical Gaussians for the core feature. The latter was typically
the brightest feature at the extreme end of a one-sided jet in most
sources (see the Appendix and Paper~VI for a discussion of core
identifications and two-sided jets in the sample). The parameters of
the Gaussian fits are listed in Table~\ref{3gaussiantable}. In some
instances, it was not possible to robustly cross-identify the same
components in a jet from one epoch to the next. Those components with
robust cross-identifications over at least 5 epochs for the purpose of
kinematics analysis are indicated in column 10 of
Table~\ref{3gaussiantable}. For the non-robust components, we note
that the assignment of the same identification number across epochs in
Table~\ref{3gaussiantable} does not necessarily indicate a reliable
cross-identification.

\input{5velocitytablestub}

We estimate errors on the component sizes to be roughly twice the
positional error, according to \cite{1999ASPC..180..301F}.  The errors
on the peak flux density values are approximately 5\% (see Appendix A
of \citealt{2002ApJ...568...99H}). Based on our previous analysis from
Paper~VI, we estimate the typical uncertainties in the Gaussian
centroid positions to be $\sim1/5$ of the FWHM beam dimensions. For
isolated bright and compact components the positional errors are
smaller by approximately a factor of two.  A more quantitative
estimate for individual components can be obtained using the scatter
of our kinematic fit residuals (columns 14 and 15 of
Table~\ref{5velocitytable}).  These residuals represent only estimates
of the uncertainty of the fits, and are likely underestimates in some
cases due to possible errors in component cross-identification and/or
a low number of epochs. Small variations in the apparent core
position, due to changes in opacity and/or newly emerging features,
can also contribute to the positional errors.  Deviations from linear
or simple accelerated motion can also increase the magnitude of the
fit residuals (see Section~\ref{accelerations}).

In Paper~VI there were eight jets which had no robust jet components.
After our analysis using the new data, six remained in this category:
0235+164, 0727$-$115, 1124$-$186, 1324+224, 1739+522, and 1741$-$038.
In the case of 0109+224 and 0742+103, we did not consider any
components to be robust in Paper~VI, due to gaps in temporal coverage.
We have subsequently obtained several closely spaced VLBA epochs and
now consider several slow-moving components in these two jets to be
robust.

In Paper~VI we listed 0048$-$097 and 1958$-$179 as having one robust
component each. However, after re-examining the original modelfits
along with the new data, we have determined that these two jets are
too compact at 15 GHz to classify any of their components as robust. 
Of the new AGN not in Paper~VI, there are six with no
robust components: 0716+332, 0946+006, 1921$-$293, 1959+650,
2023+335, and 2247$-$283. 

\subsection{Jet Kinematics Analysis}\label{kinematics}

\input{accelmotiontablestub}

We performed two sets of kinematics analyses on the robust Gaussian
jet components in our sample. The first assumed a simple
non-accelerating, two-dimensional vector fit to the component position
over time, referenced to the core component (which we presumed to be
stationary). For the components which had measurements at 10 or more 
epochs, we also performed a constant acceleration fit (as described in
Paper~VI), which yielded kinematic quantities at a reference (middle)
epoch. The results of this analysis are listed in
Table~\ref{5velocitytable} and Table~\ref{acceltable}.
Table~\ref{5velocitytable} gives the mean flux density, core
separation distance and position angle, as well as the best fit proper
motion vector and reference epoch for each robust component. For components which show significant ($\ge3\sigma$) accelerations, the
values listed in Table~\ref{5velocitytable} are from the acceleration
fit. 

\begin{figure*}
\centering
\includegraphics[angle=270,width=0.8\textwidth]{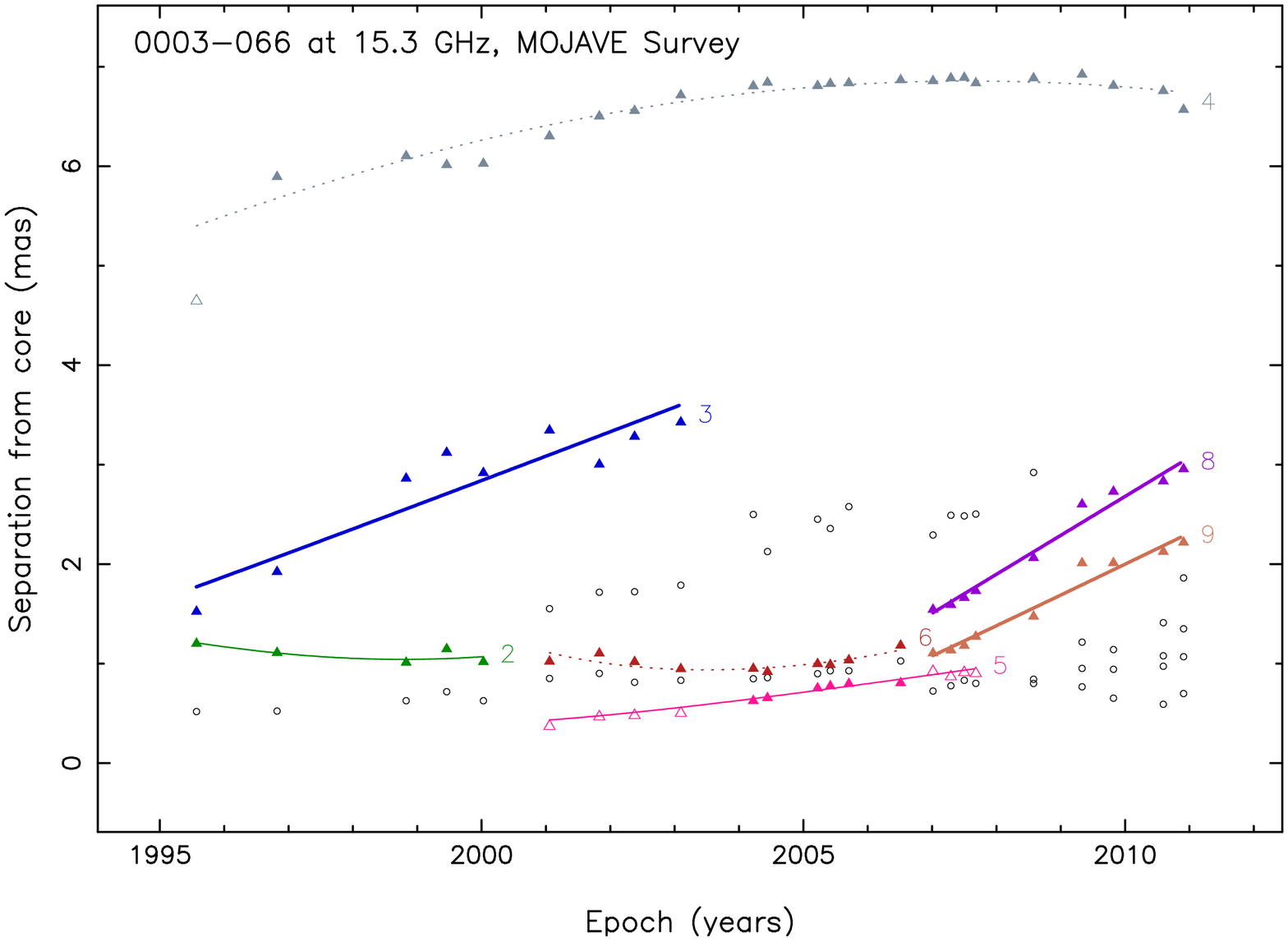}
\caption{\label{f:sepvstime}
Plot of angular separation from core versus epoch for Gaussian jet
components. Estimates for the positional errors (not plotted) are
listed in Table~\ref{5velocitytable}. The B1950 source name is given at the top left of each
panel. Colored symbols indicate robust components for which kinematic
fits were obtained (dotted and solid lines). The solid lines indicate
vector motion fits to the data points assuming no acceleration, while
the dotted lines indicate accelerated motion fits. Thick lines are
used for components whose fitted motion is along a radial direction
from the core, while the thin lines indicate non-radial
motions. Unfilled colored circles indicate individual data points that
were not used in the kinematic fits, and unfilled black circles
indicate non-robust components.  The component identification number
is indicated next to the last epoch of each robust component. (This is
a figure stub; an extended version is available online.)  }
\end{figure*}

\begin{figure*}[p]
\begin{center}
\includegraphics[angle=0,width=0.82\textwidth]{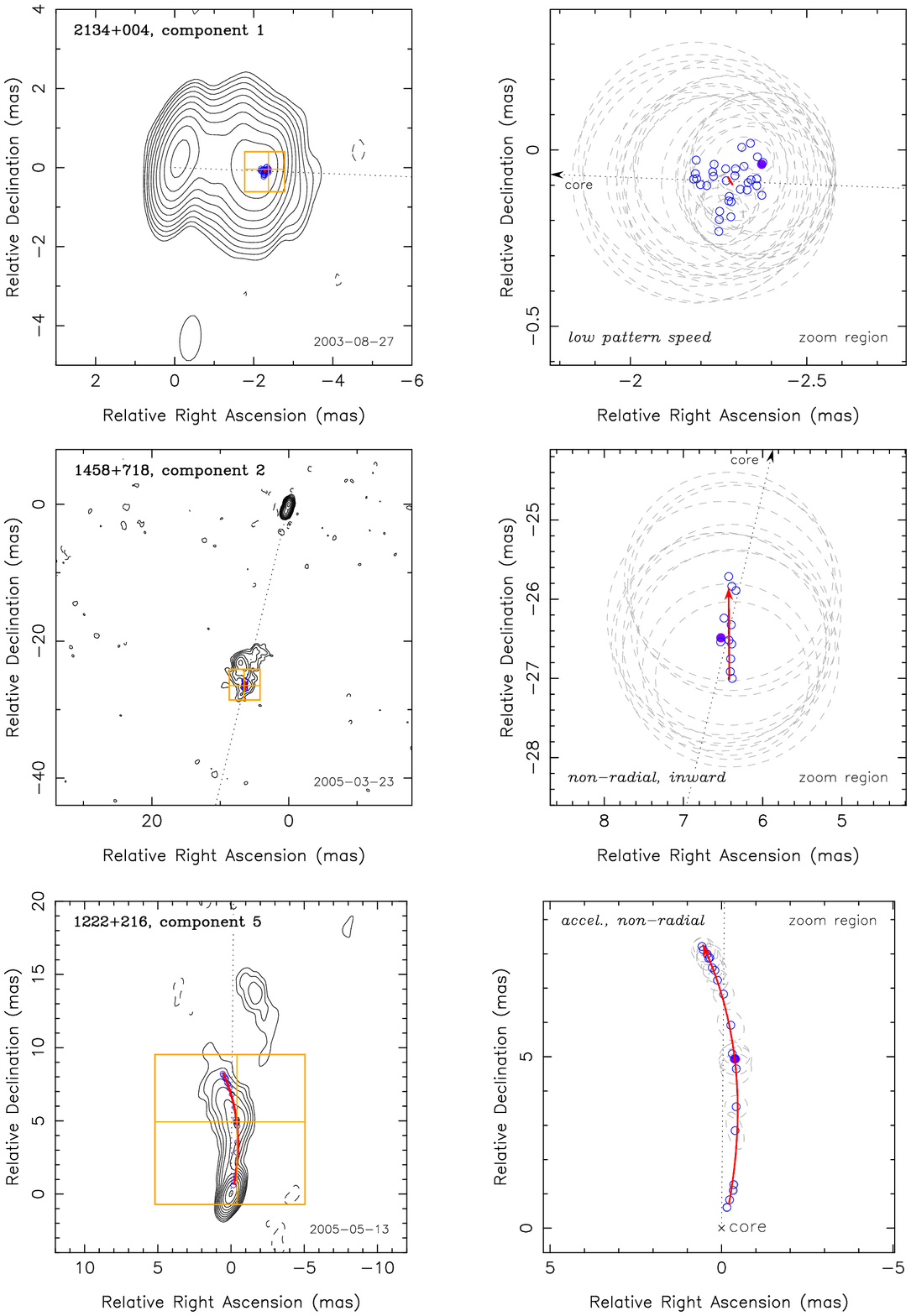}
\end{center}
\caption{\label{xyplot_a} Vector motion fits and sky position plots of
  individual robust jet components in MOJAVE AGN. Positions are relative
  to the core position. The left hand panels show a 15 GHz VLBA
  contour image of the jet at the middle epoch listed in
  Table~\ref{5velocitytable}. The orange box delimits the zoomed region
  displayed in the right hand panels. The component's position at the
  middle epoch is indicated by the orange cross-hairs. The dotted line
  connects the component with the core component and is plotted with the
  mean position angle $\langle\vartheta\rangle$
  (Table~\ref{5velocitytable}). The position at the middle epoch is
  shown by a filled violet circle while other epochs are plotted with
  unfilled blue circles. The red solid line indicates the vector (or
  accelerating) fit (see Table~\ref{5velocitytable}) to the component
  positions. The red arrows in the right hand panels indicate the
  direction of motion, and the gray dashed circles/ellipses represent
  the FWHM sizes of the individual fitted Gaussian components.
  Displayed from top to bottom in the figure are component ID = 1 in
  2134+004, ID = 2 in 1458+718, and ID=5 in 1222+216. }
\end{figure*}

\begin{figure*}
\begin{center}
\includegraphics[angle=0,width=0.82\textwidth]{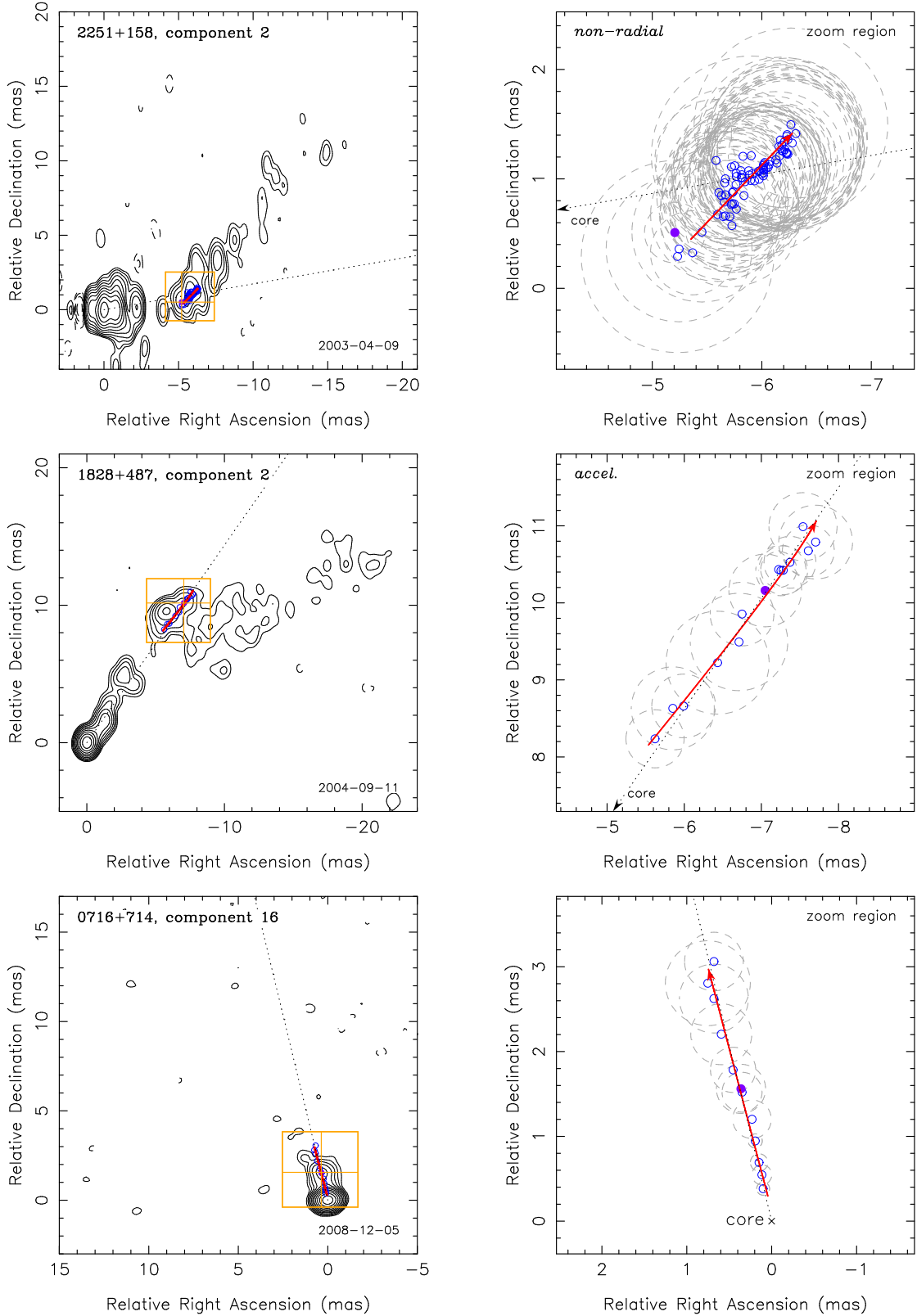}
\end{center}
\caption{\label{xyplot_b} Continued.  Displayed from top to bottom in
  the figure are component ID = 2 in 2251+158, ID = 2 in 1828+487, and
  ID=16 in 0716+714. (This is a figure stub; the complete set is
  available online.) }
\end{figure*}

Figure~\ref{f:sepvstime} shows the radial separation of the components
from the core component over time for each jet. The left hand panels
of Figures~\ref{xyplot_a} and \ref{xyplot_b} show the motions of
individual components on the sky, as well as a 15 GHz VLBA contour
image of the jet at the middle epoch listed in
Table~\ref{5velocitytable}. The orange box delimits the zoomed region
displayed in the right hand panels. Extended versions of these
figures, containing plots of all of the robust jet components, are
available online.

\begin{deluxetable}{ll} 
\tablecolumns{2} 
\tabletypesize{\scriptsize} 
\tablewidth{0pt}  
\tablecaption{\label{kinematicsummary}Summary of Kinematics Analysis Results}  
\tablehead{\colhead{Property} & \colhead {Total Number}  }
\startdata 
AGN analyzed for kinematics  & 200 \\
Jet components classified as robust  & 887 \\
Robust slow pattern speed components &\phn 38 (\phn 4\%) \\
Robust jet components with ($\ge 3\sigma$) measured speeds  & 739 (83\%)\\
Robust inward-moving ($\ge 3\sigma$ speed) components &\phn 17 (\phn 2\%) \\
Components with significant ($\ge 3\sigma$) non-radial motion & 282 (38\%) \\
Components analyzed for acceleration & 547 (62\%) \\
Components with significant acceleration & 212 (39\%) \\
Components with significant perpendicular acceleration &\phn 99 (18\%) \\
Components with significant parallel acceleration & 155 (28\%)
\enddata
\end{deluxetable} 

We examined each component with significant proper motion $(\mu \ge
3\sigma_\mu)$ for non-radial motion by comparing the mean position
angle of the component $\langle\vartheta\rangle$ with its proper
motion vector direction $\phi$.  We flagged any component for which
the angular offset $|\langle\vartheta\rangle - \phi|$ was $\ge3\sigma$
from either $0\arcdeg$ or $180\arcdeg$ as ``non-radial'', and
``inward'' if the offset was significantly greater than 90\arcdeg.  We
made these determinations using the $|\langle\vartheta\rangle - \phi|$
values from the acceleration fit for significantly accelerating
components, or from the vector fits otherwise.  Of the 739
($\ge3\sigma$) motion components classified as robust, 282 (38\%)
exhibit significant non-radial motion, while only 17 (2.3\%) are
flagged as inward (see Table~\ref{kinematicsummary}).  We discuss the
rare inward motion cases in Section~\ref{inward}.

We calculated the ejection times (defined as when the calculated core
separation equals zero) for the non-accelerating, non-inward
components by taking the average value extrapolated from the proper
motion fits in the right ascension and declination directions. We did
not compute ejection times for components which had significant vector
motion offsets (within $2\sigma$ of $15\arcdeg$ or larger), since this
would involve an extrapolation of an unknown acceleration. The errors
on the ejection times were calculated by the same method as Paper~VI;
we note that the numerical $t_{ej}$ error values tabulated in that
paper were a factor of $\sqrt{2}$ too large due to a calculation
error.

We list the parameters of the acceleration fits in
Table~\ref{acceltable}, where we have resolved the acceleration terms
$\dot\mu_\perp$ and $\dot\mu_\parallel$ in directions perpendicular
and parallel, respectively, to the mean angular velocity direction
$\phi$.  For those components with a measured redshift, a significant
proper motion $(\mu \ge 3\sigma_\mu)$, and a small uncertainty in the
radial motion offset ($|\langle\vartheta\rangle - \phi| \le 5\arcdeg$),
we calculate relative accelerations as in Paper~VII, where $ \dot\eta
= (1+z)\dot\mu/\mu$.

\section{DISCUSSION}\label{discussion}
\subsection{Apparent Inward Motions}\label{inward}

There are many scenarios under which apparent inward motions can be
produced, including: i) curved jet motions which cross the line of
sight, ii) non-stationarity of the apparent core feature, due to one
or more newly emerging features below the interferometric resolution
level, iii) a misidentification of the true stationary core with a
moving feature, iv) internal brightness changes in a large, diffuse
jet feature, and v) apparent backward pattern speeds not associated
with the flow.  In Paper~VI we identified fewer than ten individual
cases of inward motion, which represented $<2\%$ of the robust
components.  This strongly ruled out a fully random pattern speed
scenario for the component motions, in which equal numbers of inward
and outward motions would be expected.

Among all 887 robust components which we have analyzed, we find 16
statistically significant inward component motions in 10 different
jets (6 BL Lac objects, 2 quasars, and 2 radio galaxies). BL Lac jets
are statistically overrepresented in this group, given that they make
up only 20\% of the 200 AGN jets we analyzed for kinematics. We do not
find anything otherwise distinctive about these particular BL Lacs,
and the small number statistics make it impossible to draw any firm
conclusions as to the cause of this overrepresentation.  We also note
that 4 of the 10 inward component motion jets are BL Lacs with no
measured redshift. The latter is not unexpected, however, since by
definition BL Lacs have weak emission lines or featureless optical
spectra that often make it difficult or impossible to measure their
redshifts.

We previously identified three jets (1458+718, 2021+614, and 2230+114)
as having apparent inward motions in Paper~VI.  With the addition of
our new data, the jets of 2005+403, 2200+420 (BL Lac), 2201+171, and
2351+456 no longer display any statistically significant inward
motions (see Appendix for details).

The inward motions are all typically slow, with a median value of
33\muasyr, and none are faster than $\sim$ 100\muasyr.  Considering
only the AGN with a known redshift, the inward components of 1458+718
are the only ones which appear significantly superluminal, ranging
from 1.4 c to 4.6 c.  With the exception of 1458+718, 2021+614, and
2230+114, the inward motions all occur within $\sim 1$ mas of the
core, in typically the innermost component. In particular, the
innermost two jet components of two TeV-emitting BL Lacs in our
sample: 0219+428 (3C 66A) and 1219+285 (W Comae) are both
inward-moving.  The small velocities and core separations of these
moving components may indicate that the core is not a stable reference
point in these two jets. We did not find any significant inward
motions in the other 15 currently known TeV-emitting AGN
jets\footnote{http://www.tevcat.uchicago.edu} in Table~\ref{gentable}
which we analyzed.

\subsection{Parsec-Scale Jet Orientation Variations}\label{pavstime}

The tendency for the parsec-scale jets of blazars to change their
position angles on the sky with time has been solidly established via
long term VLBI studies of several individual AGN (see
\citealt{2009ASPC..402..330A} for a recent review). The exact origin
of the wobbling is not clear, although accretion disk precession,
orbital motion of the accretion system, or instabilities in the jet
flow have all been suggested. A main signature of precession is
sinusoidal variations in the jet position angle, and evidence for
this has been seen in blazars such as 3C~273
\citep{2006AA...446...71S}, 3C~345 \citep{2005AA...431..831L},
0716+714 \citep{2005AA...433..815B}, BL Lac
\citep{2003MNRAS.341..405S}, and M81 \citep{2013arXiv1301.4782M}. Other jets
have displayed monotonic position angle swings with no evidence of
periodicity (e.g., 3C~279: \citealt{2004AJ....127.3115J}; NRAO~150:
\citealt{2007AA...476L..17A}).

Until now, there has been no systematic survey of jet position angle
variations in a large blazar sample. Using our extensive MOJAVE
database, we have analyzed the innermost jet regions of 60 AGN from
our radio-selected sample for which we have obtained 20 or more VLBA
epochs over a minimum twelve year period. We excluded several
well-monitored jets in our sample for reasons of core identification
uncertainty, counterjet emission, or highly curved jet structure
within one mas of the core. We determined the innermost jet position
angle at each epoch by taking a flux density-weighted position angle
average of all clean components above 3 times the image noise level in
the annular region from 0.15 mas to 1 mas from the core. We also
explored other methods, such as using the position angle of the
innermost Gaussian modelfit component, or multiple components within a
particular distance from the core, but the derived position angles
were more influenced by choices made for the fitted Gaussians (e.g.,
elliptical versus circular, and total number of Gaussians in the inner
jet region). Our position angle measurement method is subject to
errors associated with wandering of the core component position due to
changes in opacity or the emergence of new components below our
resolution level. These are likely small, however, since they would
create correlated apparent motions in components located downstream,
which we have not detected in our data.

\begin{figure*}
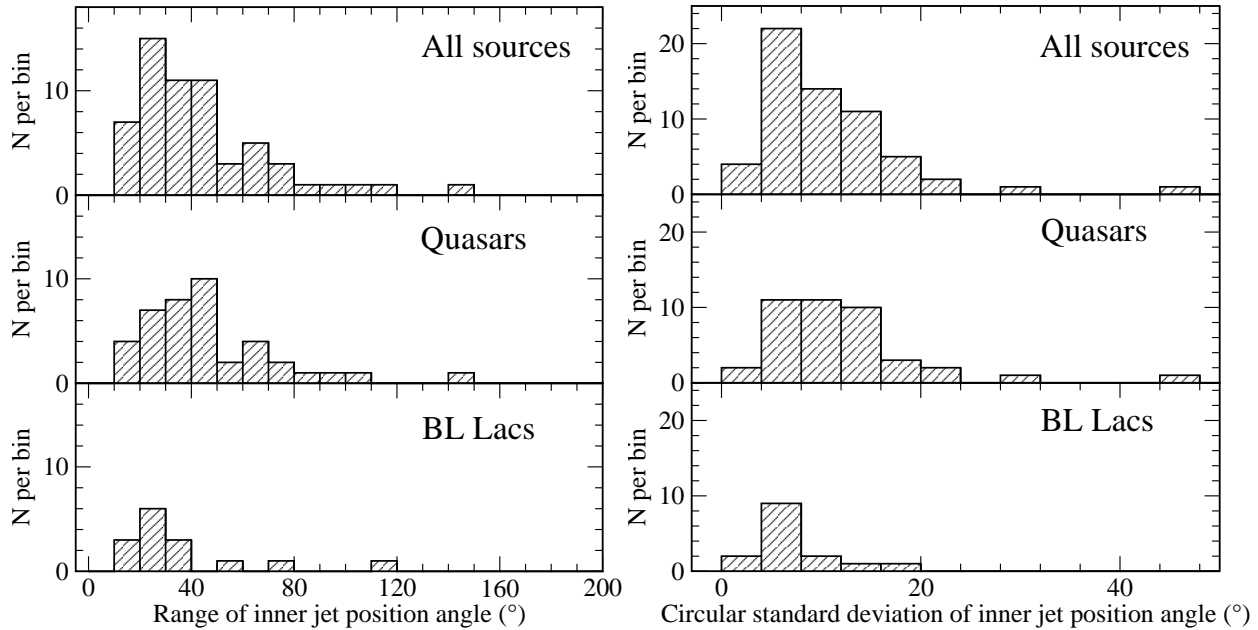

\centering
\includegraphics[trim=0.1cm 0cm 0cm 0cm,clip,width=0.45\textwidth]{parange_optclass.eps}
\includegraphics[trim=0.1cm 0cm 0cm 0cm,clip,width=0.46\textwidth]{pasigma_optclass.eps}
\caption{\label{parange} Distribution of innermost jet position angle
  variation over time for 60 individual MOJAVE radio-selected jets
  observed over at least 12 years.  Left panel: Distribution of
  position angle range (maximum minus minimum value). Right panel:
  Distribution of circular standard deviation. In each panel, the top
  plot shows the distribution of position angle range (maximum minus
  minimum) for all jets, while the middle and bottom plots show the
  distributions for quasars and BL Lacs, respectively.}
\end{figure*}

We find that the innermost jet position angles vary considerably over
the 12 to 16 year intervals covered by our data, with ranges up to
$150 \arcdeg$ in some jets (Fig.~\ref{parange}). The typical circular
standard deviation in position angle is $\sim 10\arcdeg$. The quasars
and BL Lac objects differ significantly in their ranges, according to
Kolmogorov-Smirnov tests on their range ($p_\mathrm{null}$ = 1.1\%)
and standard deviation ($p_\mathrm{null}$ = 1.3\%) distributions. It
is unclear whether the smaller overall variations we see in the BL Lac
innermost jet position angles is an intrinsic effect, or because they
are oriented at slightly larger angles to the line of sight than
quasars.

Some jets (e.g., NRAO~150 = 0355+508) show a very wide range of inner
jet position angle and Gaussian component position angles, likely
because the viewing angle to the inner jet lies within the opening
angle of the (presumably conical) outflow. In other cases, such as
3C~273 (1226+023), the jet is tranversely resolved into multiple
features, some of which are moving along different position angles at
nearly the same radial distance from the core. Finally, the
distributions shown in Fig.~\ref{parange} are not peaked in the first
bin, implying that it is common in blazars for features to emerge at
different position angles.  Since stacked-epoch VLBA images (Paper~V)
often show a smooth conical jet intensity profile in highly variable
ejection angle jets such as 1308+326, the simplest interpretation is
that individual emerging features do not fill the entire cross-section
of the flow.  Instead, features are ejected within a finite width
``ejection cone'', which becomes apparent only in a stacked-epoch
image. The apparent opening angle of this cone is exaggerated by
projection effects by a factor of $\sin{\theta}$. Thus for typical
blazar jet viewing angles of $\theta \lesssim 5\arcdeg$, the intrinsic
ejection cone full opening angle is likely $\sim 0.5\arcdeg$ to $\sim
2\arcdeg$, based on the position angle range distribution in
Fig.~\ref{parange}.  AGN jets can therefore appear ``bent'' in a
single-epoch, limited dynamic range VLBI image, whereas in reality
what is visible is just the portion of the jet that is currently
experiencing enhanced synchrotron emission, due to the passage of
several very bright features.  These features may represent flow
instabilities driven at the nozzle (e.g.,
\citealt{2011IAUS..275...41H}), which vary within the jet over time
and in turn influence the kinematics downstream.

\begin{figure*}[p]
\centering
\epsscale{1.15}
\plotone{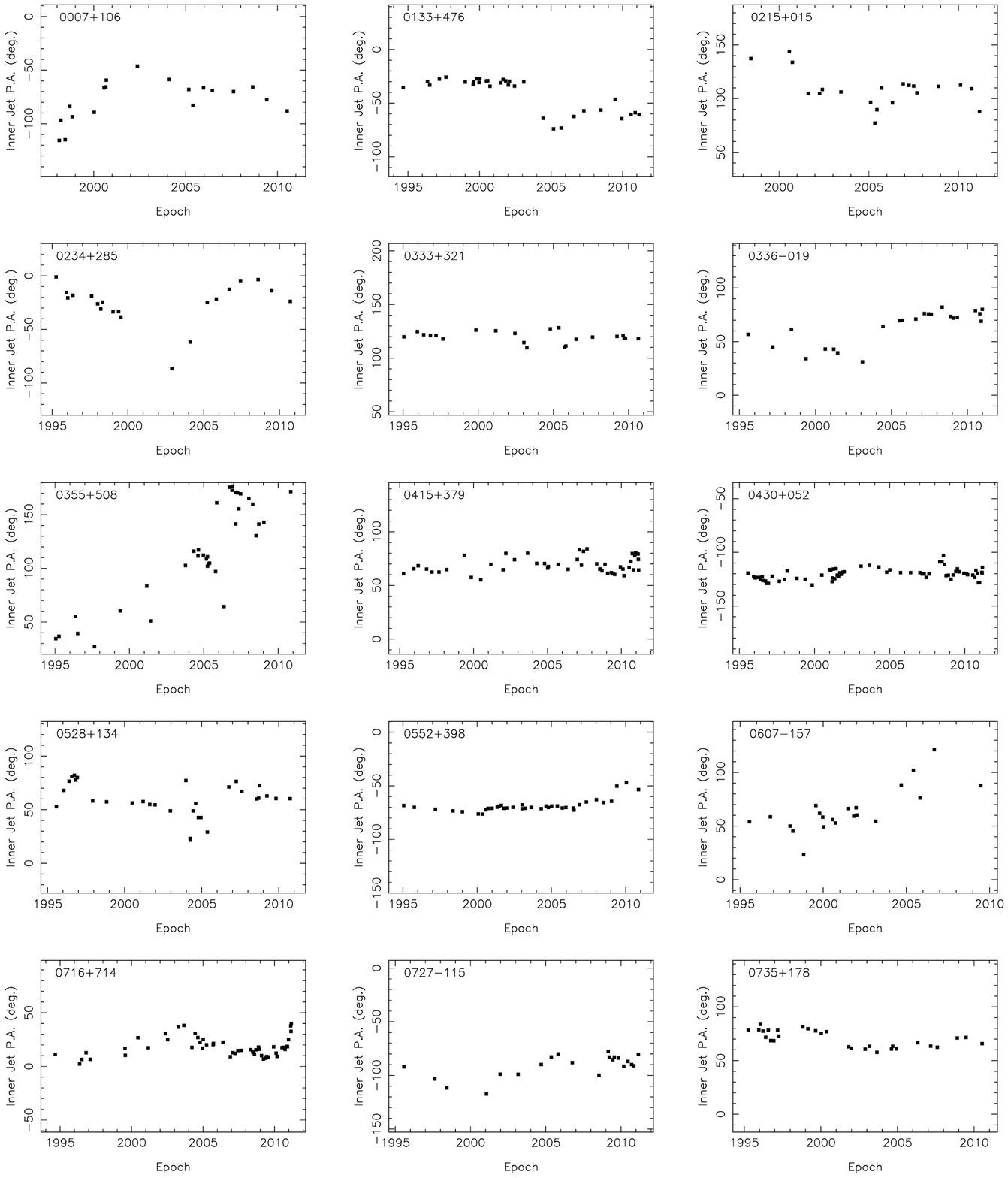}
\caption{\label{pavstimeall} Innermost jet position angle versus time
  plots for 60 selected MOJAVE jets having 20 or more VLBA epochs over
  at least 12 years.  (This is a figure stub; an extended version is available online.)}
\end{figure*}

In Figure~\ref{pavstimeall} we plot the innermost jet position angle
derived from the clean components for each jet versus time on the same
vertical scale. The individual jets show a variety of behavioral
patterns, which we have classified into four general categories. There
are 14 jets which display a monotonic trend, 5 which show a back and
forth trend, 12 which show more than one cycle of back and forth
motion (oscillatory), and 29 with no discernible trend.  Within these
categories, 11 jets exhibited one or more abrupt jumps in position
angle, caused by a new feature emerging from the core with a
significantly different trajectory than previously ejected features.

\begin{deluxetable}{llc} 
\tablecolumns{3} 
\tabletypesize{\scriptsize} 
\tablewidth{0pt}  
\tablecaption{\label{regressionfittable} Linear Regression Fits to Monotonic Innermost Jet
  Position Angle Trends}  
\tablehead{\colhead {Source} & \colhead {Slope } &  \colhead {Fit}  \\
 \colhead{Name} & \colhead {(deg. $\mathrm{y^{-1}}$}) & \colhead {Range} }
\startdata 
0355+508& $\phn 9.8 \pm 1$& 1995.1--2010.8 (all data)\cr
0607$-$157& $\phn 4.6 \pm 0.9$& 1995.6--2009.5 (all data)\cr
0748+126& $\phn 2.2 \pm 0.4$ & 1997.0--2011.0 data only\cr
0851+202& $-2.6 \pm 0.1$& 1995.3--2010.0 data only\cr
0906+015& $-0.7 \pm 0.2 $& 1997.6--2011.2 (all data)\cr
1226+023& $-1.9 \pm 0.2 $& 1995.6--2010.8 (all data)\cr
1253$-$055& $-1.5 \pm 0.1$& 1995.6--2007.0 data only\cr
1633+382& $\phn 2.2 \pm 0.2$& 1995.1--2011.1 (all data)\cr
1807+698& $\phn 0.4 \pm 0.1 $& 1996.0--2011.1 data only\cr
2005+403& $\phn 3.8 \pm 0.7$& 1995.1--2011.1 (all data)\cr
2200+520 & $-0.8 \pm 0.2$& 1995.3--2011.3  (all data)\cr
2223$-$052& $\phn 2.8 \pm 0.5$& 1995.6--2010.7 (all data)\cr
2230+114& $\phn 2.1 \pm 0.4 $&    1998.5--2010.8 data only\cr
2251+158& $-2.0 \pm 0.1$& 1995.6--2011.3 (all data)
\enddata
\end{deluxetable}

We performed linear regression fits on the jets showing monotonic
trends, and have tabulated the fitted slopes in
Table~\ref{regressionfittable}. In some cases, we had to omit segments
of the data from the fit due to abrupt jumps in position angle (see
column 3 of Table~\ref{regressionfittable}).  The most rapidly varying
jet, at $9.8 \pm 1 \arcdeg \; \mathrm{y^{-1}}$, is 0355+508 (NRAO
150). Our measurement is consistent with the swings of up to $11
\arcdeg \; \mathrm{y^{-1}}$ seen previously in this jet by
\citealt{2007AA...476L..17A} in VLBA images at 43 and 86 GHz. The
other monotonically changing jets in our sample show swings of
typically a few degrees per year. In the particular case of 0851+202
(OJ~287), \cite{2012ApJ...747...63A} found erratic variations in the
innermost jet position angle at 43 GHz, with amplitudes $< 40\arcdeg$
and timescales, $\lesssim 2$ years, as well as an abrupt jump in
position angle in late 2004. Our observations at 15 GHz, on the other
hand, show a monotonic swing in position angle of $-2.6\pm 0.1 \arcdeg
\; \mathrm{y^{-1}}$ from 1995 until the end of 2010, when the jet
underwent a sudden large jump in position angle. Earlier VLBI
observations, as tabulated by \cite{2012MNRAS.421.1861V}, show that
this monotonic swing began in the early 1990s. Superimposed on the
longterm trend are small amplitude wiggles with timescales of several
years (Fig.~\ref{pavstimeall}). The 2004 jump seen at 43 GHz is not
present in the 15 GHz data, and the 2010 jump seen at 15 GHz is not
evident in the 43 GHz data.  This likely reflects the different
angular resolution of the two datasets.  \cite{2012ApJ...747...63A}
used the position angle of Gaussian components out to $\sim 0.4$ mas
from the core, with no flux density weighting, whereas at 15 GHz we
used a flux density-weighted average of clean components in the inner
0.15 to 1 mas region. It is also possible that OJ~287 emits very
short-lived, bright features visible at 43 GHz which rapidly fade
before they can be resolved in our 15 GHz images.



\begin{figure*}[t]
\centering
\epsscale{1.15}
\plotone{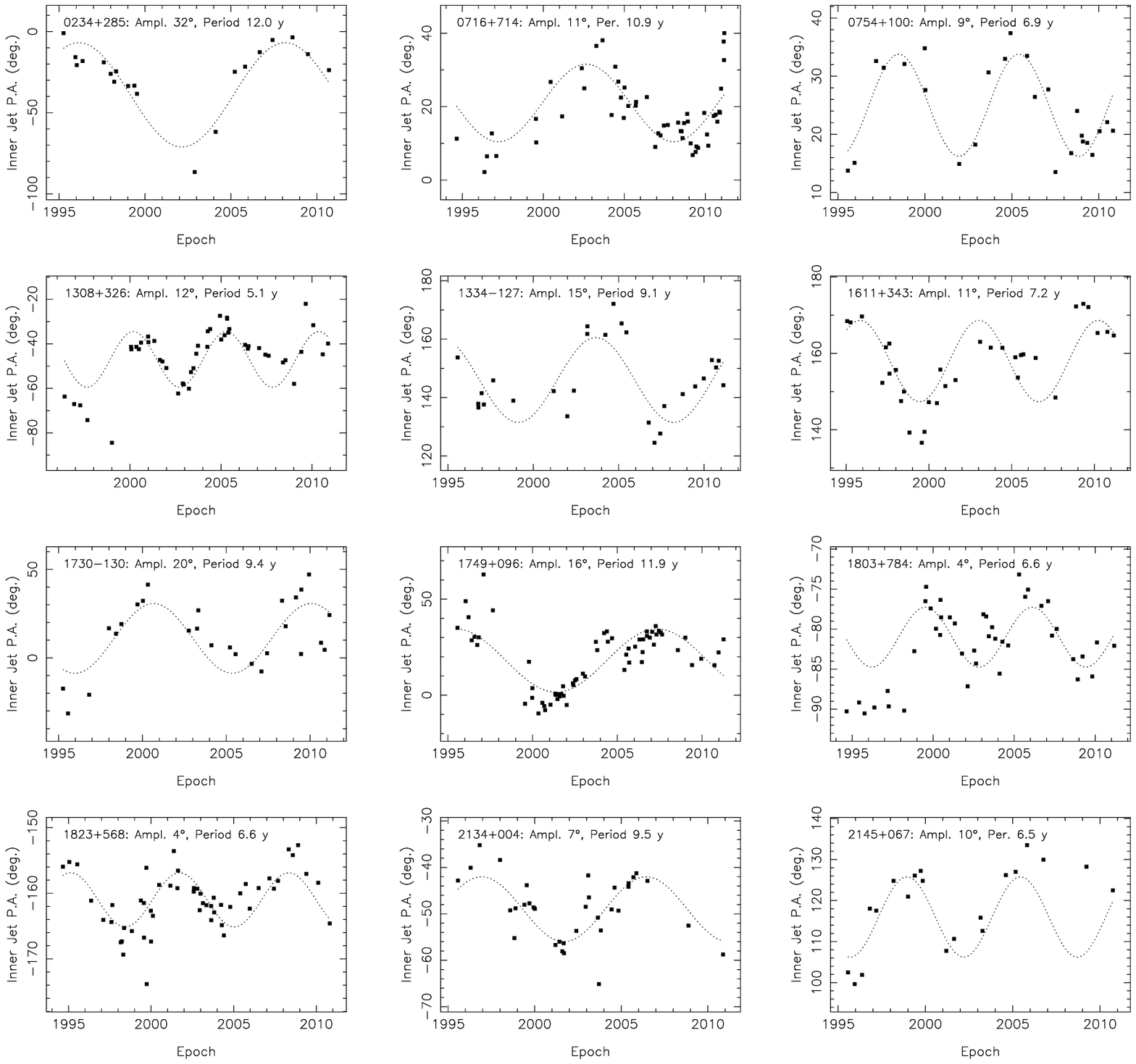}
\caption{\label{allsinefits}Innermost jet position angle versus time
  plots for 12 selected MOJAVE jets displaying oscillatory trends. The
  dotted lines represent the best sine curve fits to the data, based
  on the peak period in the Lomb-Scargle periodogram. }
\end{figure*}

We looked for evidence of periodicity in all 60 jets using
Lomb-Scargle periodograms, which are well-suited for unequally sampled
time-series data \citep{1976ApSS..39..447L,1982ApJ...263..835S}. With
the exception of 0234+285 and 2145+067, the 12 jets shown in
Figure~\ref{allsinefits} have significant ($\ge2\sigma$) peaks at the
periods indicated on each sub-panel, which range from 5 to 12 y.  The
dashed lines represent the best sinusoid fits to the data for the
indicated Lomb-Scargle period. For these fits we allowed the
amplitude, mean, and phase to vary in order to find the best absolute
$\chi^2$ value. Assuming a typical Gaussian normal error of $2\arcdeg$
for the position angle measurements (based on the linear regression
fit residuals to the monotonic trend sources), the best
reduced-$\chi^2$ fit values are 1.5 for 0716+714, and 2.5 for
1823+568. We cannot reliably establish periodicity in any jet,
however, due the lack of sufficient cycles in the data, and the fact
that many of the fits have significant residuals (e.g., 1308+326,
1803+784, 2145+067), suggesting that the behavior is more complex than
a single sinusoidal variation.

\subsection{Dispersion of Apparent Speeds}\label{speeddispersion}

A longstanding question in the study of AGN jet kinematics is whether
the bright features in a given jet all tend to propagate at a
characteristic speed which represents the true flow. In Paper~VI we
found that roughly 20\% of the jets had one or more features which
moved significantly slower than the other features in the jet. These
slow pattern speed features could be the result of either stationary
shocks in the flow, or jet bending across the line of sight. We have
repeated this analysis for the current data set, using more stringent
criteria.

We first tabulated a maximum speed for each jet by considering the
component with the fastest $\ge3\sigma$ speed. If no component in the
jet had a $\ge3\sigma$ speed, we lowered the criterion to $>2\sigma$. In
the case of 0355$+$508, 1329$-$049, and 1520$+$319, which had no
$\ge2\sigma$ components, we used the component speed with the smallest
measured error.  We dropped 16 jets from the analysis since they had no
robust components with which to measure a maximum speed. We 
flagged components in Table~\ref{3gaussiantable} as ``Slow Pattern
Speed'' (SPS) if they had i) no statistically significant
acceleration, ii) a speed less than 20\muasyr, and iii) a speed at
least 10 times slower than the fastest component in the jet. We found 38
such components in 29 different jets (14 quasars, 10 BL Lacs, and 4
radio galaxies).

A significantly higher fraction of BL Lacs and radio galaxies in our
sample contain SPS components ($\sim 25$\%) as compared to
quasar jets (10\%). We detect no significant differences in the redshift
distributions of SPS and non-SPS jets, but their median VLBA 15 GHz
luminosity distributions are different at the 98\% confidence level
according to a Kolmogorov-Smirnov test. Of the 37 jets with radio luminosity above
$10^{28}$ W/Hz, only one has a SPS component (the quasar 2134$+$004).
This is consistent with numerical simulations (e.g.,
\citealt{1994ApJ...436L.119D,1999ApJ...516..729R}) which show that
more highly relativistic jets exhibit fewer compact internal structures and
less overall instability.


In order to examine the dispersion of speeds within individual jets in
more detail, we calculated a normalized statistic $D = (\mu_\mathrm{max} - \mu_\mathrm{min}) / (\mu_\mathrm{max} + \mu_\mathrm{min})$
for all 75 jets which had at least five robust
components, as well as a median speed and rms dispersion value.  Over
half of these jets have $D$ values above 0.8, indicating that in most
cases, the overall range of apparent speed within a jet is comparable
to its maximum speed. Removing the SPS components from the analysis
did not change this general result.

\begin{figure*}
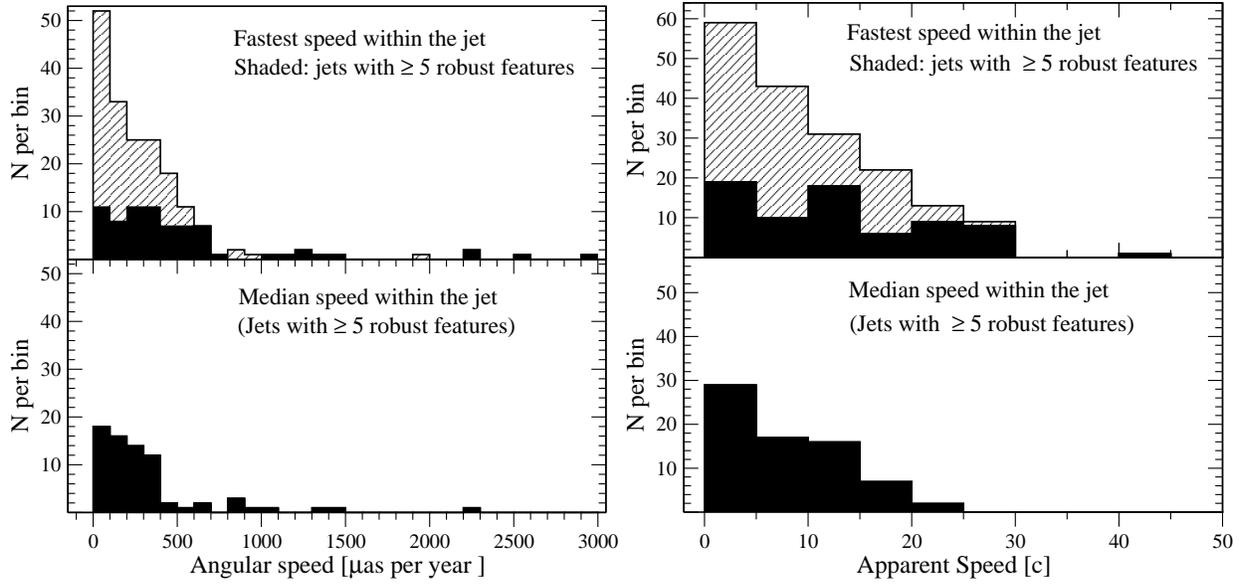

\centering
\includegraphics[trim=0cm 0cm 0cm 0cm,clip,width=0.45\textwidth]{angspeedhist.eps}
\includegraphics[trim=0cm 0cm 0cm 0cm,clip,width=0.45\textwidth]{betaspeedhist.eps}
\caption{\label{angspeedhist}Distributions of fastest and median
  speed for jets in our sample with measured redshifts. The left-hand
  panels show angular speeds in \muasyr, while the right-hand panels are in units
  of the speed of light. The shaded histograms are for jets having at
  least 5 robust components. }
\end{figure*}

Although we find there can be a large range of apparent speed within a
jet, the speeds usually cluster around a median value, indicating they
are not random. The median rms dispersion of speed within a jet is 2.8
$c$. This is significantly less than the overall dispersion of the
median values (6.7$c$).  Furthermore, the distributions of maximum and
median speed in the sample are not uniformly distributed in either
angular or spatial measure, but are instead peaked at low values
(Fig.~\ref{angspeedhist}).

\begin{figure*}
\centering
\includegraphics[trim=0cm 0cm 0cm 0cm,clip,width=1.0\textwidth]{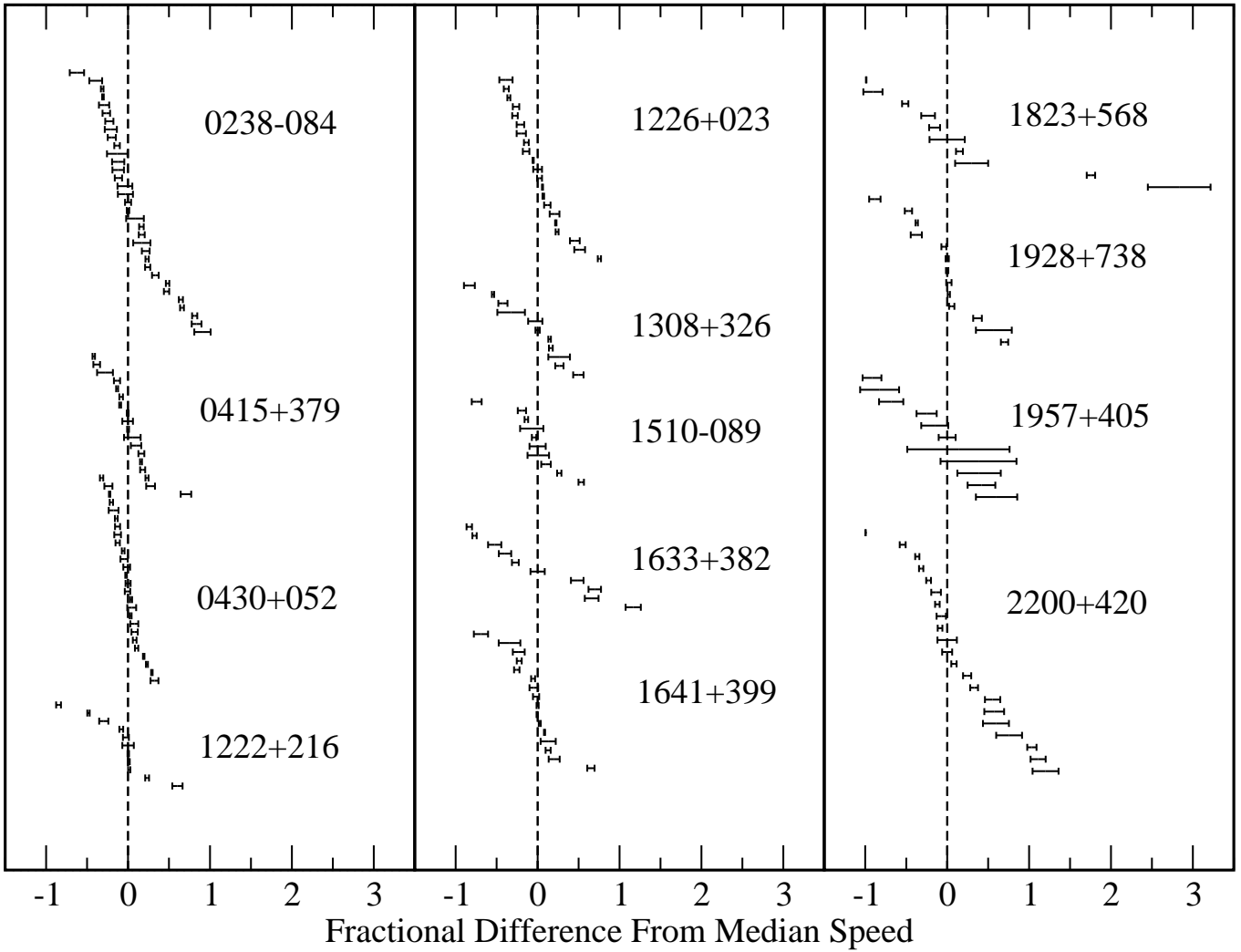}
\caption{\label{ladderplot}Distribution of apparent speeds within
  jets having at least 10 robust components. The components in
  each jet are arranged vertically from slowest to fastest, and the speeds are normalized
  with respect to the median speed of each jet. The fractional
  difference is defined as $(\mu - \mu_\mathrm{median}) / \mu_\mathrm{median}$. }
\end{figure*}

\begin{figure}[t]
\centering
\includegraphics[trim=0cm 0cm 0cm 0cm,clip,width=0.45\textwidth]{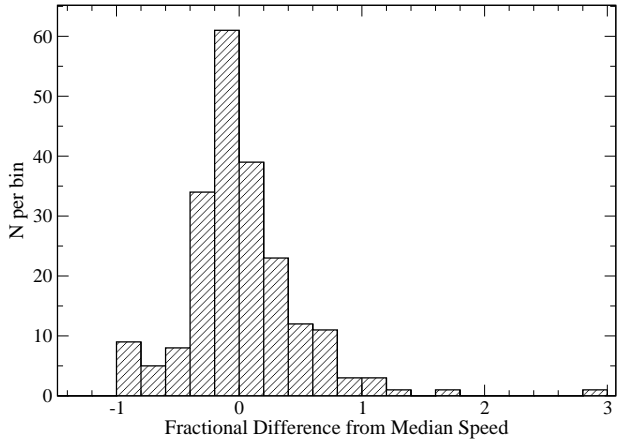}
\caption{\label{ladderhist}Overall normalized speed distribution
  within jets having at least 10 robust components.  The fractional
  difference is defined as $(\mu - \mu_\mathrm{median}) / \mu_\mathrm{median}$. }
\end{figure}

In Figure~\ref{ladderplot} we plot a representation of the apparent
speed distribution within each of the 13 jets in the sample which have
at least 10 robust components.  For each component, we plot $(\mu -
\mu_\mathrm{median}) / \mu_\mathrm{median}$, the fractional difference
with respect to the median speed in the jet. The components in each
jet are arranged vertically from slowest speed at the top to fastest
speed at the bottom. We use this format instead of a binned
representation in order to display the error bars. The binned
distribution for the combined set of 13 jets is shown in
Figure~\ref{ladderhist}.

There is a range of behavior seen among these jets, with some (e.g.,
0238$-$084, 0430+052, 1641+399, and 2200+420) having a near-Gaussian
speed distribution, and others such as 1633+382 and 1823+568 ejecting
features uniformly distributed over a wide range of speeds.
Nevertheless, the combined distribution in Fig.~\ref{ladderhist} shows
a near-Gaussian shape, although with significant kurtosis (more
sharply peaked than Gaussian, according to an Anscombe-Glynn test).
The distribution has no significant skewness if the outlier at 2.8 is
omitted (D'Agostino test). One might expect an asymmetric distribution
(positively skewed) if some fraction of the features travel at the
true flow speed and none of them exceed it, but we do not see evidence
of this in the data. Since we can typically identify more robust
components in the jets for which have the longest monitoring
intervals, we might also expect to see a positive trend between the
maximum speed of a jet and the number of robust components, but this
is not evident in our data. Our findings thus suggest that in general,
high ejection rate blazar jets tend to eject features with apparent
speeds which cluster about a speed that is characteristic to each
individual jet.

A dispersion of apparent speeds within a jet might be expected in the
context of our model presented in Section~\ref{pavstime}, in which
features do not fill the entire jet cross section, and emerge along a
radial streamline within a conical outflow. Evidence of thin
ribbon-like instabilities has been seen in the transversely resolved
jet 3C~273 \citep{2001Sci...294..128L} and the quasar S5~0836+710
\citep{2012ApJ...749...55P}. The latter authors conclude that the
ridgeline in VLBI images may correspond to a helically twisted
pressure maximum within the jet, which slowly varies in position on
decadal timescales. In the case of 3C~273, polarimetric VLBA rotation
measure maps made at different epochs by \cite{2005ApJ...626L..73Z},
\cite{2008ApJ...675...79A}, and \cite{2012AJ....144..105H} indicate
that different parts of the jet are being illuminated at different
times. Similar behavior has also been reported in the broad line radio
galaxy 3C~120 by \cite{2011ApJ...733...11G}.

\begin{figure}
\centering
\includegraphics[angle=-90,width=0.47\textwidth]{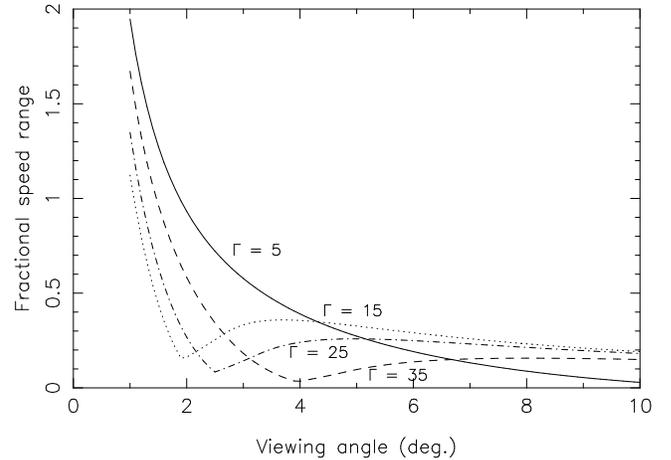}
\caption{\label{speedrange}Expected range of apparent speed for features
  emitted within a conical jet of one degree half-angle, plotted
  against on-axis viewing angle. The y-axis values represent
  fractional difference from the on-axis apparent speed. The solid
  curve is for a jet with Lorentz factor 5, the dotted curve has
  Lorentz factor 15, the dot-dashed curve Lorentz factor 25, and the
  dashed curve Lorentz factor 35.  }
\end{figure}

In our proposed scenario we would expect a range of apparent viewing
angle roughly equal to the full opening angle of the ejection cone,
and a corresponding range in apparent speed. In
Figure~\ref{speedrange} we plot the range of expected apparent speed
for conical jets with a 2$\arcdeg$ full opening angle versus on-axis
viewing angle. The y-axis values are normalized with respect to the
on-axis apparent jet speed, i.e., a value of one indicates an apparent
speed range as large as the on-axis apparent speed. The curves
represent different bulk flow Lorentz factors, and have minima at the
critical angle $(1/\Gamma)$ at which superluminal motion is maximum.
Given the typically small fractional errors on our apparent speed
measurements, the only cases where the predicted range approaches the
FWHM of the data distribution ($\sim 1$, see Fig.~\ref{ladderhist})
are for exceedingly aligned jets where the viewing angle is comparable
to the jet opening angle.  Since the full opening angles of powerful
blazar jets are typically less than two degrees
\citep{2005AJ....130.1418J,2009AA...507L..33P}, and most blazar jets
in a flux density-limited sample will be viewed at approximately 1/2
the critical superluminal motion angle \citep{1994ApJ...430..467V},
examples of jets viewed inside their opening angle should be rare.
Such jets should be characterized by wide apparent opening angles and
a large range in the sky position angles of ejected features (i.e.,
the outlier jets in Fig.~\ref{parange}). We find no indications,
however, that either of these two quantities are correlated with the
range of apparent speed for the individual jets in our sample. We
therefore conclude that the spread in apparent speed we see in
individual jets cannot be wholly attributed to a spread in streamline
viewing angle.  The moving features must also have an intrinsic range
of bulk Lorentz factor and/or pattern speed.

\begin{figure*}
\centering
\includegraphics[angle=-90,width=0.98\textwidth]{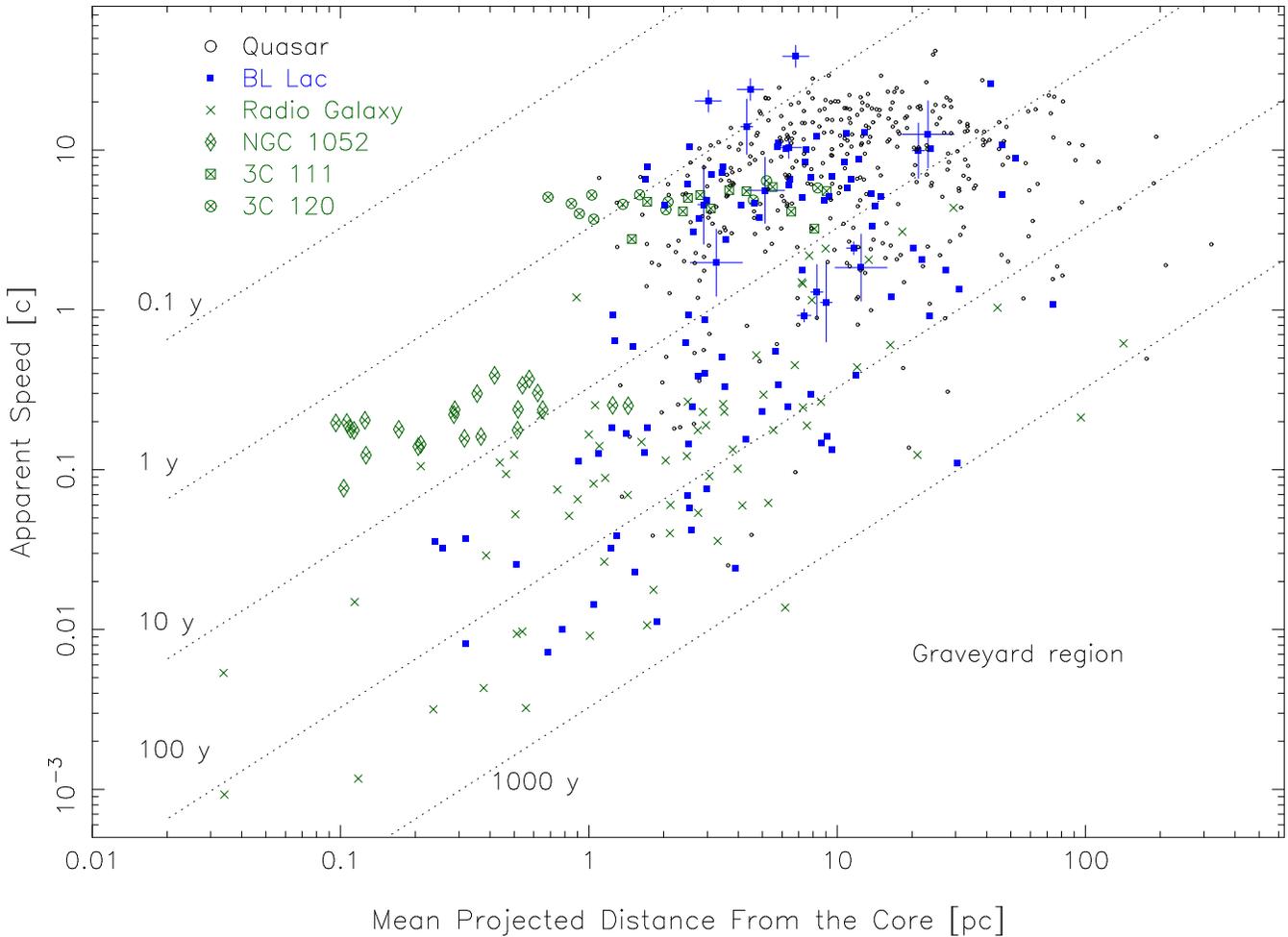}
\caption{\label{beta_vs_pc}Apparent speed versus mean
  projected distance from the core in parsecs for all robust jet
  components (excluding non-radial and inward components).  The green
  crosses denote components in radio galaxies, the black circles
  quasars, and the blue squares BL Lacs. The jet components of two
  broad-lined radio galaxies (3C~111 and 3C~120), as well as the
  gigahertz-peaked spectrum radio galaxy NGC~1052 are highlighted with
  distinct symbols. Error bars have been omitted for the purposes of
  clarity.  For components in jets which do not have a spectroscopic
  redshift, we include range bars which correspond to their known
  redshift constraints (see Appendix).  The dotted lines indicate
  lines of constant age for components that have moved steadily
  outward over the time period indicated. Newly emergent fast
  components rapidly evolve out of the top left region of the plot,
  while no components are found in the bottom right graveyard region
  since they will likely have faded well below the MOJAVE imaging
  sensitivity level. }
\end{figure*}

\subsection{Trends With Apparent Speed}\label{speedcorr}

\subsubsection{Speed versus distance down the jet}\label{speedvsdistance}

In our analysis of MOJAVE and VLBA archive images from 1994--2007
(Papers VI and VII) we found that close to the base of the jet,
features tended to show increasing rather than decreasing apparent
speed. This suggests that AGN jet flows are still being organized on
pc-scales, a favored possible site for high-energy photon production.
In order to test for this trend in our current dataset, we have
plotted the apparent speed of all robust components versus their mean
projected distance from the core in Figure~\ref{beta_vs_pc}. We have
omitted components with non-radial or inward motions, and have
included lines of constant age which assume steady radial motion over
the indicated time period. Since the components in our sample have a
wide range of ages, this would tend to suppress any artificial trend
of higher mean distance for faster components. However, no components
are found in the upper left corner of the plot, since these fast
components will have higher mean core distance values and quickly
evolve towards the right hand side of the diagram. Also, sufficient
time must pass to gather sufficient epochs for an apparent speed
measurement in our survey.  Components are also absent from the lower right
``graveyard'' region since these evolved components have likely
undergone considerable adiabatic expansion and synchrotron energy
losses, thereby dropping their flux densities below the threshold for
which we can robustly measure their centroids and speeds.  There is
also a notable deficit of components above $\sim 0.4$ c within $\sim
1$ pc of the core, which would be even more pronounced without the
inclusion of the numerous components associated with the two-sided
jets of the gigahertz-peaked spectrum galaxy NGC~1052. The two broad
line radio galaxies 3C~111 and 3C~120 also occupy a distinct region
among the radio galaxy data plotted in Figure~\ref{beta_vs_pc}. 

The overall distribution of the components in the plot indicates a
positive correlation of speed with core distance for radio galaxies
and BL Lac objects, even after partialling out redshift. This trend
needs to be confirmed, however, using a larger, complete AGN sample
which extends below 0.35 Jy.  It is not possible to assess the
existence of a trend for the quasars due to the observational
selection effects described above.  Because of the low space density
of bright quasars at low redshift and the fixed angular resolution
limit of our survey, there is likely an artificial deficit of quasar
jet components in the region below 1 c and within $\sim 1$ pc of the
core, precisely where data are needed to test for a possible trend.

As we reported previously in Paper~VI, the BL Lac jets have, on
average, slower component speeds than the quasars.  There are,
however, many BL Lac jet components that overlap with the region
populated by quasar jet components in Figure~\ref{beta_vs_pc}.
\cite{2012MNRAS.420.2899G} discuss the possibility of quasars
masquerading as low-equivalent line width BL Lacs due to the swamping
of their emission lines by highly Doppler boosted optical synchrotron
emission. We note one of the jet components with the fastest angular
speed in our sample, ID = 16 in the BL Lac object 0716+714, is a
distinct outlier in Figure~\ref{beta_vs_pc}, with recent constraints
on its redshift by \cite{2013ApJ...764...57D} implying an apparent
speed of up to $43.6 \pm 1.3\; c$. This AGN is highly variable at all
wavelengths (e.g., \citealt{1996AJ....111.2187W,2013ApJ...768...40L},
and has a Doppler factor of at least 20
\citep{2013AA...552A..11R}, making it a strong candidate for a
masquerading BL Lac object.

\begin{figure*}
\centering
\includegraphics[angle=-90,width=0.98\textwidth]{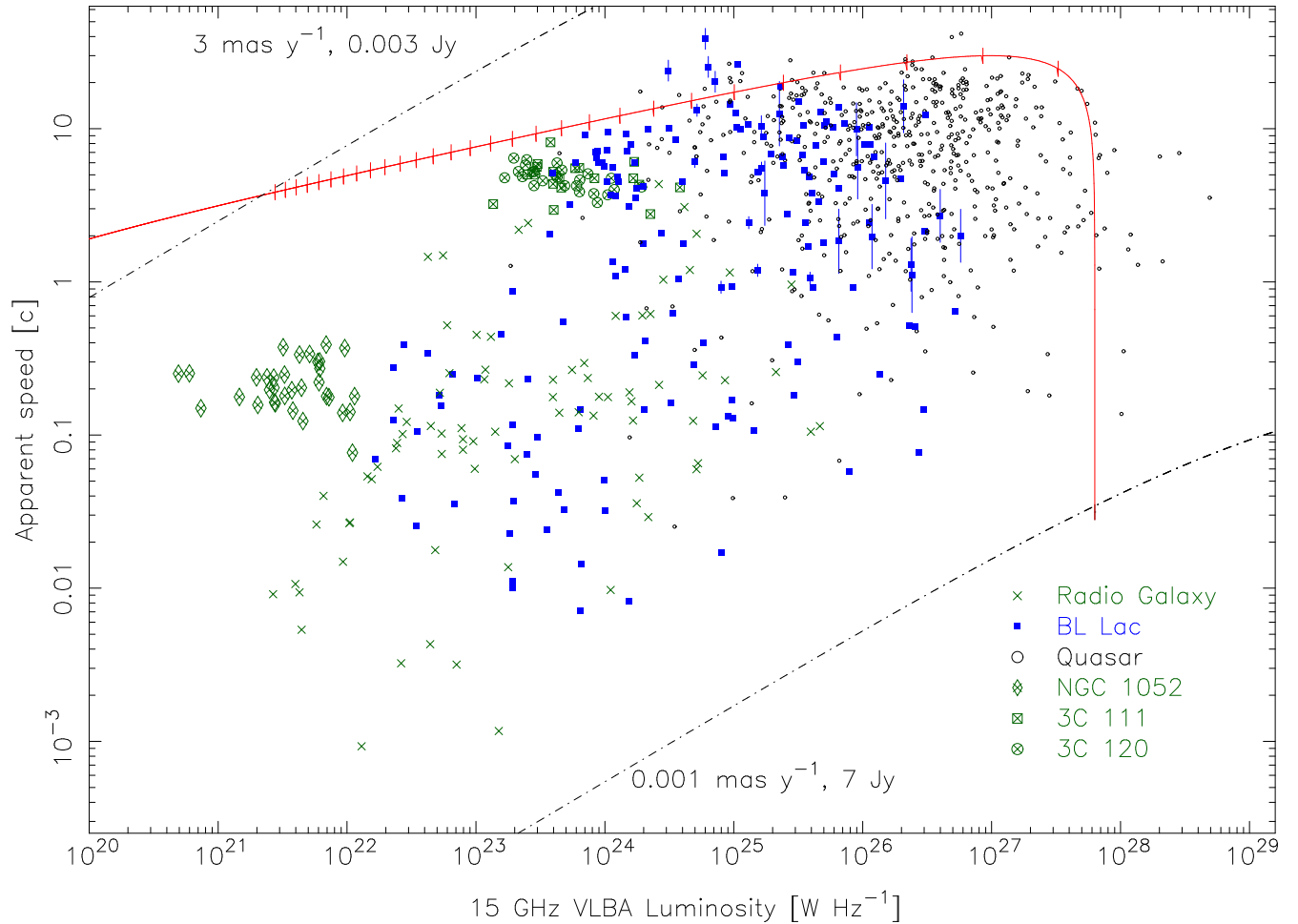}
\caption{\label{betalumcpts}Apparent speed versus 15 GHz
  luminosity for all robust jet components (excluding inward
  components).  The green crosses denote components in radio galaxies,
  the black circles quasars, and the blue squares BL Lacs. Error bars
  have been omitted for purposes of clarity.  For components in jets
  which do not have a spectroscopic redshift, we include vertical
  range bars which correspond to their known redshift constraints (see
  Appendix).  The red curve shows the locus of possible locations for
  a jet component having intrinsic luminosity $10^{23} \; \mathrm{W\;
    Hz^{-1}}$, Doppler boost $\delta^{2.7}$ and a Lorentz factor
  $\Gamma = 35$, viewed at different angles to the line of sight. The
  tick marks are drawn at one degree intervals, ranging from a viewing
  angle of 30$\arcdeg$ on the left to $1\arcdeg$ on the right. The
  dot-dashed lines correspond to observational limits; the regions in
  the top left and bottom right corners of the plot are not sampled by
  the survey.}
\end{figure*}

\subsubsection{Speed versus luminosity}

A plot of apparent speed versus luminosity for the robust jet
components (Fig.~\ref{betalumcpts}) reveals a distinct deficit of fast
components at low luminosities.  The apparent upper envelope to this
distribution has been discussed in several studies
(\citealt{1995PNAS...9211385V,1994ApJ...430..467V, LM97,
  2007ApJ...658..232C}, Paper~I, Paper~VI), and roughly matches the
locus of points for a jet of fixed intrinsic luminosity and bulk
Lorentz factor, oriented at different angles to the line of sight (for
$L > 10^{23}\; \mathrm{W \; Hz^{-1}}$). The parametric red curve in
Figure~\ref{betalumcpts}) is drawn for a jet with $\Gamma = 35$,
$L_\mathrm{int} = 10^{23} \; \mathrm{W\; Hz^{-1}}$, and Doppler boost
$\delta^{2.7}$.  This curve is representative only, since jets in the
population have a range of Lorentz factors and intrinsic luminosities,
and thus a family of such curves exist, as described by
\cite{2007ApJ...658..232C}. The upper left and lower right regions of
the plot (delimited by the dot-dashed lines) are not sampled by our
survey. The deficit region located below the red curve and below
$10^{23}\; \mathrm{W \; Hz^{-1}}$ is partly due to the incompleteness
and relatively high flux density cutoff of our low-luminosity sample.
The overall fall-off in the upper edge of the distribution from $10^{26}$ $\mathrm{W\;
  Hz^{-1}}$ to $10^{23}$ $\mathrm{W\; Hz^{-1}}$ is not the result of
survey bias or selection effects, however, and reflects both the
existence of a maximum jet Lorentz factor in the parent population
($\sim 40$), and an intrinsic correlation between flow speed and
luminosity in AGN jets. The relatively sharp edge to the upper
envelope in Figure~\ref{betalumcpts}, as well as the relatively
unchanging distribution of apparent speed in luminosity bins above
$10^{26} \;\mathrm{W\; Hz^{-1}}$ imply that intrinsically powerful AGN
jets have a wide range of Lorentz factors up to $\sim 40$, while
intrinsically weak jets are only mildly relativistic.

\subsection{Accelerating Components}\label{accelerations}

Theoretical models of the acceleration and collimation of blazar jets
indicate that the strong magnetic fields associated with the putative
supermassive black hole/accretion disk system play a key role in the
initial acceleration and collimation of the jet
\citep[e.g.,][]{2001Sci...291...84M}.  While some models indicate that
this process is largely complete with the conversion of Poynting flux
to flow energy on sub-pc scales \citep[e.g.,][]{2005ApJ...625...72S},
there may still be significant magnetic \citep{2004ApJ...605..656V} or
hydrodynamic acceleration which extends to parsec or decaparsec scales.

\begin{figure*}[p]
\centering
\includegraphics[angle=0,width=0.85\textwidth]{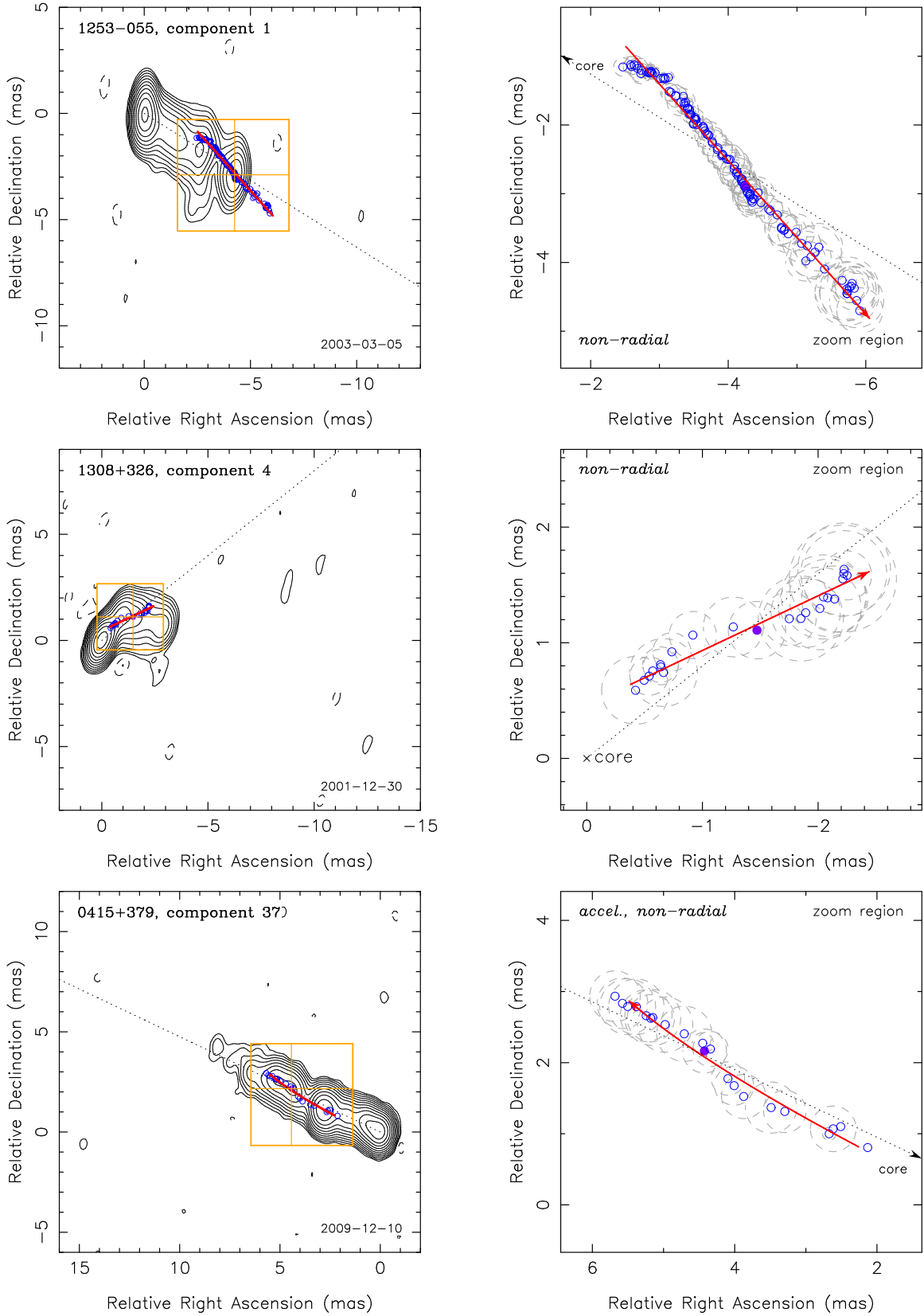}
\caption{\label{impulsive_accel} Vector motion fits and
  sky position plots of several robust jet components which show
  impulsive or multiple accelerations that are poorly fit by a simple
  constant acceleration model.}
\end{figure*}

In our previous analysis of MOJAVE data from 1994--2007 (Paper~VII),
we found that accelerated motions with respect to the mean apparent
velocity vector $\vec\beta_\mathrm{obs}$ were common, with significant
parallel accelerations seen in roughly one third of our sample, and
significant perpendicular accelerations in about one fifth of our
sample. Due to the limited number of available epochs, we were only able to
analyze 311 of the 526 moving components for possible accelerated
motion.  Using our new data spanning up to 2011 May 1, we have
performed the same analysis on 547 of the 887 moving components in
Table~\ref{5velocitytable} which have least 10 epochs. We summarize our
overall acceleration results in Table~\ref{acceltable}. Our incidence
rates of parallel accelerations (28\%) and perpendicular accelerations
(18\%) are consistent with those reported in Paper~VII. Nearly 40\% of
the components analyzed show significant acceleration of either type.
A substantial number of components showed no significant acceleration,
but had non-radial motion vectors. If we assume that these components
accelerated some time prior to our monitoring interval, the fraction
of moving components showing accelerations rises to 70\%. This is in
stark contrast to the kinematics of features in stellar (Herbig-Haro)
jets, which are well described by ballistic models (e.g.,
\citealt{2013AA...551A...5E}).

We note that our acceleration fits assume a simple two-dimensional
parameterization in which the acceleration components in the R.A. and
declination directions are constant over time. Impulsive or multiple
changes in a component's velocity are therefore not well-accommodated
in our model. In Figure~\ref{impulsive_accel} we show some examples of
component motions which are poorly represented by a simple constant
acceleration fit. We discuss this issue, as well as the detailed
acceleration properties of the full sample in a future paper in this
series.

\section{SUMMARY}\label{conclusions}

We present 1753 new multi-epoch VLBA images of 259 jets which are
drawn from a complete radio-selected AGN sample above 1.5 Jy, a
complete $\gamma$-ray selected sample \citep{2011ApJ...742...27L}, and
a representative low-luminosity radio jet sample. The latter is drawn
from the VLBA Calibrator Survey and consists of 43 AGN with 8 GHz
VLBA flux density greater than 0.35 Jy, $z \le 0.3$, and J2000
declination $> -30\arcdeg$.

We have combined these new VLBA data with existing MOJAVE observations
previously presented in Paper~V, and have analyzed the kinematics of
200 parsec-scale AGN jets. The data span 1994 Aug 31 to 2011 May 1,
and were modelled with Gaussian components in the visibility plane. We
obtained vector motion fits to 887 robust components which were
positively cross-identified over at least 5 VLBA epochs. We also
measured the acceleration properties of 557 components which had at
least 10 VLBA epochs. Our main findings are as follows:

1. Nearly all of the 60 most heavily observed jets show significant
changes in their innermost position angle ($20\arcdeg$ to
$150\arcdeg$) over a 12 to 16 year monitoring interval, with BL Lac
jets showing smaller variations than quasars.  The observed range
corresponds to intrinsic (de-projected) variations of $\sim
0.5\arcdeg$ to $\sim 2\arcdeg$.  Roughly half of the heavily observed
jets display a trend of innermost jet position angle with time.  In
the case of 12 jets, there is some evidence of oscillatory behavior,
but the fitted periods (5 to 12 y) are too long compared to the length
of the dataset to firmly establish periodicity.  These periods are
very short compared to expected precession timescales from warped
accretion disk - black hole interactions associated with the
Bardeen-Petterson effect \citep{1998ApJ...506L..97N}. Rather, we favor
a conical jet model in which emerging features do not fill the entire
cross-section of the flow. What is typically visible in a
single-epoch, limited dynamic range VLBA image as a series of bright
features may actually be the lit up portions of thin ribbon-like
structures embedded within a broader conical outflow. These portions
have been energized by a passing planar disturbance which originated
in the VLBI core. As discussed by \cite{2012ApJ...749...55P}, the
ribbon structures may arise from helical pressure maxima within the
jet, and slowly vary in position on decadal timescales.

2. We examined the distribution of speeds within 75 jets which had at
least 5 robust components.  Within a particular jet, the speeds of the
features usually cluster around a characteristic value, however, the
the range of apparent speed among the components is comparable to the
maximum speed measured within the jet. This is not due uniquely to
unusually slow pattern speed components (defined as having a speed
less than $20 \;\mathrm{mas\;y^{-1}}$ and at least 10 times slower
than the fastest component in the jet), which comprise only 4\% of all
the components studied and are less prevalent in quasar jets. It is
also too large to be solely due to differences in ejection angle. We
conclude that the dispersion is at least partially due to an intrinsic
distribution of bulk Lorentz factor and/or pattern speed.

3. Apparent inward motions are rare, with only 2\% of the components
having apparent velocity vectors greater than 90$\arcdeg$ from the
outward jet direction. They occur more frequently in BL Lac jets.  All
of the detected inward motions are slow ($\lesssim 0.1\;
\mathrm{mas\;y^{-1}}$), and nearly all occur within 1 mas of the core
component.

4. We confirm a previously established upper envelope to the
distribution of speed versus beamed luminosity for moving jet
components.  Below $10^{26} \; \mathrm{W\;Hz^{-1}}$ there is a
fall-off in maximum speed with decreasing 15 GHz radio luminosity. The
relatively sharp edge to the upper envelope, as well as the relatively
unchanging distribution of apparent speed in luminosity bins above
$10^{26} \;\mathrm{W\; Hz^{-1}}$ imply that intrinsically powerful AGN
jets have a wide range of Lorentz factors up to $\sim 40$, while
intrinsically weak jets are only mildly relativistic.

5. We find a trend of increasing apparent speed with distance down
the jet for radio galaxies and BL Lac objects. The existence of a
trend could not be evaluated for quasars due to unavoidable
observational selection biases, as described in Section~\ref{speedvsdistance}.

6. Accelerations are very common among the moving jet components. Of
the 739 components with statistically significant ($\ge3\sigma$) speeds,
38\% exhibited significant non-radial motion, implying non-ballistic
trajectories. We analyzed 547 components with at least 10 epochs, and
found 28\% to have significant accelerations parallel to the velocity
vector, and 18\% to have significant perpendicular accelerations.
Nearly 40\% showed significant acceleration of either type, and an
additional 30\% had non-radial motion vectors.

The MOJAVE program continues to gather VLBA data on the AGN jets in
the samples presented here. Future papers in this series will discuss
their polarization evolution, detailed acceleration characteristics,
and connections between jet kinematics and other blazar properties
such as $\gamma$-ray emission, flux variability, and SED synchrotron peak
frequency.

\acknowledgments
The MOJAVE project was supported under NASA-{\it Fermi} grants
NNX08AV67G and 11-Fermi11-0019. ER was partially supported by the
Spanish MINECO projects AYA2009-13036-C02-02 and AYA2012-38491-C02-01
and by the Generalitat Valenciana project PROMETEO/2009/104, as well
as by the COST MP0905 action `Black Holes in a Violent Universe'. MFA
and HDA were supported by NASA-{\it Fermi} GI grants NNX09AU16G,
NNX10AP16G and NNX11AO13G and NSF grant AST-0607523. YYK was supported
by the Russian Foundation for Basic Research (projects 12-02-33101,
13-02-12103), 
Research Program OFN-17 of the Division of Physics, Russian Academy of Sciences,
and the Dynasty Foundation. ABP
was supported by the ``Non-stationary processes in the Universe''
Program of the Presidium of the Russian Academy of Sciences. DCH was
supported by NSF grant NSF AST-0707693. The National Radio Astronomy
Observatory is a facility of the National Science Foundation operated
under cooperative agreement by Associated Universities, Inc. This work
made use of the Swinburne University of Technology software correlator
\citep{2011PASP..123..275D}, developed as part of the Australian Major
National Research Facilities Programme and operated under licence. The
area-proportional Venn diagram was produced using the eulerAPE
package.


\bibliographystyle{apj}
\bibliography{lister}

\appendix
\section{Notes on Individual Sources}

0048$-$097:  No distinct features could be reliably tracked in this
very compact jet. 

0111+021: The second and third closest components to the core in this
nearby BL Lac object (z = 0.047) have significantly inward but slow
($< 40$\muasyr) motions over six VLBA epochs spanning five years.

0219+428 (3C 66A):  No reliable spectroscopic redshift exists for this
BL Lac object. \cite{2013ApJ...766...35F} have set limits of $0.3347 \le z
\le 0.41$ based on intergalactic absorption features. 

0235+164: The jet structure was too compact at 15 GHz to reliably
measure any robust components. 

0238$-$084 (NGC 1052): Multi-frequency VLBA observations of this
two-sided jet by \cite{2003AA...401..113V} and
\cite{2004AA...426..481K} indicate that the core feature is obscured
at 15 GHz by strong free-free absorption associated with a
circumnuclear torus. In order to obtain the component positions at
each epoch, the position of a virtual core was found using a least
squares minimization method, as described by
\cite{2003AA...401..113V}. The core component entries in
Table~\ref{3gaussiantable} for this source are therefore left blank.

0300+470: No reliable redshift exists for this BL Lac object.
\cite{2013ApJ...764..135S} derive a statistical upper limit of $z <
1.63$ based on the absence of intergalactic absorption features. With
the addition of a new epoch in 2008, a component (ID = 2) first
identified in Paper~VI in this BL Lac jet now has a statistically
significant inward motion.

0430+289: No reliable redshift exists for this BL Lac object.
\cite{2013ApJ...764..135S} derive a statistical upper limit of $z <
1.66$ based on the absence of intergalactic absorption features, and
\cite{2010ApJ...712...14M} obtain $z > 0.48$ based on optical
non-detection of the host galaxy.

0506+056: This AGN is a {\it Fermi} LAT-detected, high-spectral peaked BL
Lac object with unknown redshift. \cite{2010ApJ...712...14M} obtain $z
> 0.38$ based on optical non-detection of the host galaxy, and 
\cite{2012AA...538A..26R} set an upper limit $z < 1.24$ based on the
photometric redshift technique.  A weak component on the western edge of
this jet (ID = 3) shows a statistically significant inward motion of
$100\pm 14$ \muasyr. This corresponds to an apparent speed between 2.3 and
6.1 c, given the redshift limits. 

0716+332: There were no components strong enough or sufficiently isolated to be
considered robust in this jet. 

0716+714: A spectroscopic redshift has yet to be obtained for this BL
Lac object. \cite{2008AA...487L..29N} published a value of $0.31 \pm
0.08$ based on a host galaxy magnitude estimate, while
\cite{2013ApJ...764...57D} have constrained the redshift to the range
$0.2315 < z < 0.322$ using intervening absorption systems. We note
that if the redshift is at the upper limit of this range, 0716+714
becomes one of the fastest jets in our sample ($43.6 \pm 1.3 \; c$).
For all redshift values in the possible range, the jet is a distinct
outlier in the $\beta_\mathrm{app}$ versus luminosity plot
(Figure~\ref{betalumcpts}).

0727$-$115: We did not identify any components in this jet as sufficiently
robust to track over time.

0814+425: No reliable spectroscopic redshift exists for this BL Lac
object. \cite{2005ApJ...635..173S} found $z > 0.75$ based on a lower
limit to the host galaxy magnitude, and \cite{2013ApJ...764..135S}
derive a statistical upper limit of $z < 2.47$ based on the absence of
intergalactic absorption features. The innermost component of this jet
showed significant inward motion between 1995 and 2006.  We did not
previously consider this component to be robust in Paper~VI due to
confusion arising from an emerging new component in 2007 (now identified
as ID = 6).

0946+006: There was too much positional scatter in the brightest
downstream jet component (ID = 2) to consider it as robust over the short
two-year long VLBA coverage of this AGN. 

0954+556 (4C +55.17): This unusual quasar is largely resolved by the VLBA at 15
GHz, yet has strong and variable $\gamma$-ray emission.
\cite{2011ApJ...738..148M} have suggested that it may be a young
radio source. We were not able to identify any robust components in
this jet. 

1011+496: This TeV-detected BL Lac object (1ES 1011+496;
\citealt{2007ApJ...667L..21A}) has a jet component moving at $1.8 \pm
0.4$ $c$, making it a rare example of a superluminal high-spectral
peaked blazar.

1124$-$186: The jet structure was too compact at 15 GHz to reliably
measure any robust components. 

1228+126 (M87): We confirm the slow speeds we measured in Paper~VI for
this nearby radio galaxy. The closest component to the core in the
main jet (ID = 6) has a very slow pattern speed ($3.5 \pm 4
\muasyr$), and the fastest component (ID = 4) has significant
non-radial motion at $0.026 \pm 0.003$ $c$. These speeds are
significantly slower than those measured in the HST-1 feature ($0.6
\pm 0.3$ $c$; \citealt{2010AA...515A..38C}), located more than 80 pc
further down the jet.  \cite{2007ApJ...660..200L} found speeds of 0.25
to 0.4 $c$ in the region 2-4 mas downstream from the core, based on
five VLBI epochs at 22 and 43 GHz obtained between 1999 and 2004.

1324+224: The jet structure was too compact at 15 GHz to reliably
measure any robust components. 

1424+240: No reliable spectroscopic redshift exists for this BL Lac
object. Although \cite{2010ApJ...712...14M} obtained $z = 0.23$ based
on its host galaxy magnitude, \cite{2013ApJ...764..135S} subsequently
set a firm lower limit of $z > 0.6035$ based on intergalactic
absorption features.

1458+718: We confirm the apparent inward motion reported in Paper~VI
of two components in a complex emission region located $\sim$ 25 mas
south of the core in this compact steep spectrum quasar. With our new
data we have found one additional component in this complex (ID = 2)
that is also moving inward, in a non-radial direction.  The inward
speeds of the three components range from $\sim$ 1.4 $c$ to 4.6 $c$.


1509+054: The radio structure of this AGN consists of three bright
components. We identify the radio core as the middle component, based on
the spectral index map of T. Hovatta et al. (in prep.).

1637+826: This nearby Seyfert 2 radio galaxy (also known as NGC 6251,
at $z =0.024$) contains five outward-moving components with $\ge3\sigma$
speeds, all of which are below 0.15 $c$. The innermost component (ID =
8), however, has a small but significant inward motion of $50 \pm 10$
\muasyr (0.08 $\pm$ 0.015 $c$).

1739+522: The jet structure was too compact at 15 GHz to reliably
measure any robust components.

1741$-$038: The jet structure was too compact at 15 GHz to reliably
measure any robust components. 

1823+568: This BL Lac jet at $z = 0.664$ has a very fast component
speed: $26.2 \pm 2.6$ $c$, as compared with the next fastest BL Lac
component ($15.1 \pm 0.4$ $c$ for 0851+202). It has been classified as
a quasar by \cite{VV12}, and as a powerful FR II jet by
\cite{1993MNRAS.264..298M} based on its extended radio emission. It
therefore may be an intermediate BL Lac/quasar type object.

1921$-$293: The low declination and north-south orientation of this jet
made it impossible to robustly track any of its bright features over time.

1958$-$179: The jet structure was too compact at 15 GHz to reliably
measure any robust components. 

1959+650: The jet structure was too weak and compact at 15 GHz to reliably
measure any robust components. 

2005+403: In Paper VI we reported a possible inward motion of a
component very close to the core (ID = 6), however, our new data
indicate that there is too much complex sub-structure in this region
to reliably determine robust component positions.

2021+614: We find the two outermost components (ID = 1 and 2) to have
inward motions, albeit with very slow speeds (7.2 and 15.2 \muasyr,
respectively). The fitted motion vector for component 1 is
non-radial.  The core component location in this jet remains uncertain
(see Paper~VI). 


2023+335: The radio structure in this low galactic latitude quasar 
($-2.4^\circ$) is too strongly affected by interstellar scattering to
permit the tracking of robust jet features \citep{2013arXiv1305.6005P}.

2200+420 (BL Lacertae): The innermost component (ID = 7) of this jet
showed very little motion and an uncertain vector motion direction in
our Paper~VI analysis. With the addition of many new epochs since
2007, we now find the component to have a very slow but significant
motion of $3.8 \pm 0.6 \muasyr$ ($0.017 \pm 0.003$ $c$).

2201+171: Our previous analysis in Paper~VI suggested inward motion
for component 3, but subsequent data revealed that its fitted position
after 2007 was likely affected by a new rapidly outward-moving
component (ID = 6). As a result, we have categorized component 3 as
non-robust.

2230+114 (CTA 102): In Paper~VI we reported a single jet component at $\sim$ 6 mas
(ID = 4) as having significant inward motion, which is re-affirmed with
the most recent data.

2247$-$283: All of the fitted jet features in this source were weaker
than 100 mJy, and none could be reliably tracked over the 5 available
epochs.

2351+456: After reporting an inward-moving component (ID = 2) in
Paper~VI, we subsequently re-performed the model fits to all epochs,
including the new data, and now find no statistically significant
motion for this component ($\mu = 40 \pm 14$ \muasyr).

\end{document}

%% file: gentablestub.tex
\begin{deluxetable*}{llllccc} 
\tablecolumns{7} 
\tabletypesize{\scriptsize} 
\tablewidth{0pt}  
\tablecaption{\label{gentable} General Properties of AGNs in the Combined Samples}  
\tablehead{\colhead{B1950} & \colhead {Alias} &\colhead{2FGL Assoc.} &  
\colhead{z} &  \colhead{Ref.} & \colhead{Opt.}& \colhead{Sample} \\ 
\colhead{(1)} & \colhead{(2)} & \colhead{(3)} & \colhead{(4)} & \colhead{(5)} & 
 \colhead{(6)} & \colhead{(7)}} 
\startdata 
0003+380 & S4 0003+38& J0006.1+3821 & 0.229& \cite{1994AAS..103..349S} & Q& L \\ 
0003$-$066 & NRAO 005&\n & 0.3467& \cite{2005PASA...22..277J} & B& R \\ 
0007+106 & III Zw 2&\n & 0.0893& \cite{1970ApJ...160..405S} & G& R,L \\ 
0010+405 & 4C +40.01&\n & 0.256& \cite{1992ApJS...81....1T} & Q& L \\ 
0015$-$054 & PMN J0017$-$0512& J0017.6$-$0510 & 0.226& \cite{2012ApJ...748...49S} & Q& G,L \\ 
0016+731 & S5 0016+73&\n & 1.781& \cite{1986AJ.....91..494L} & Q& R \\ 
0048$-$097 & PKS 0048$-$09& J0050.6$-$0929 & 0.635& \cite{2012AA...543A.116L} & B& G,R \\ 
0055+300 & NGC 315&\n & 0.0165& \cite{1999ApJS..121..287H} & G& L \\ 
0059+581 & TXS 0059+581& J0102.7+5827 & 0.644& \cite{2005ApJ...626...95S} & Q& R \\ 
0106+013 & 4C +01.02& J0108.6+0135 & 2.099& \cite{1995AJ....109.1498H} & Q& G,R \\ 
0109+224 & S2 0109+22& J0112.1+2245 & 0.265& \cite{2012ApJ...748...49S} & B& G,R \\ 
0109+351 & B2 0109+35&\n & 0.450& \cite{1996MNRAS.282.1274H} & Q& R \\ 
0110+318 & 4C +31.03& J0112.8+3208 & 0.603& \cite{1976ApJS...31..143W} & Q& G \\ 
0111+021 & UGC 00773&\n & 0.047& \cite{1976ApJS...31..143W} & B& L \\ 
0116$-$219 & OC -228& J0118.8$-$2142 & 1.165& \cite{1983MNRAS.205..793W} & Q& G
\enddata 

\tablenotetext{a}{Known TeV emitter (http://tevcat.uchicago.edu).}
\tablecomments{Columns are as follows: 
(1) B1950 name, 
(2) other name,
(3) 2FGL catalog name,
(4) redshift,
(5) literature reference for redshift and optical classification,
(6) optical classification, where B = BL Lac, Q=Quasar, G = Radio galaxy, N = Narrow Line Seyfert 1, and U = Unidentified,
(7) Sample membership, where G=1FM $\gamma$-ray selected sample, R = MOJAVE 1.5 Jy sample, L = low-luminosity sample.
}

\end{deluxetable*}

%% file: maptablestub.tex
\begin{deluxetable*}{lllccrrrccc} 
\tablecolumns{11} 
\tabletypesize{\scriptsize} 
\tablewidth{0pt}  
\tablecaption{\label{maptable}Summary of 15 GHz Image Parameters}  
\tablehead{ &  & \colhead{VLBA} &\colhead{Freq.} & 
\colhead{$\mathrm{B_{maj}}$} &\colhead{$\mathrm{B_{min}}$} & \colhead{$\mathrm{B_{pa}}$} &  
\colhead{$\mathrm{I_{tot}}$} &  \colhead{rms}  &  \colhead{$\mathrm{I_{base}}$} & \colhead{Fig.} \\ 
\colhead{Source} & \colhead {Epoch} & \colhead{Code} &\colhead{(GHz)} & 
\colhead{(mas)} &\colhead{(mas)} & \colhead{(\arcdeg)} &  
\colhead{(Jy)} & \colhead{(mJy bm$^{-1}$)}  &  \colhead{(mJy bm$^{-1}$)} & \colhead{Num.} \\ 
\colhead{(1)} & \colhead{(2)} & \colhead{(3)} & \colhead{(4)} &  
\colhead{(5)} & \colhead{(6)} & \colhead{(7)} & \colhead{(8)} & \colhead{(9)}& \colhead{(10)} } 
\startdata 
0003+380  & 2006 Mar 9 & BL137B\tablenotemark{a} & 15.4 & 1.01 & 0.73 & 18 & 0.649 & 0.4 & 1.3 & \ref{images}.1   \\ 
  & 2006 Dec 1 & BL137L\tablenotemark{a} & 15.4 & 0.85 & 0.58 & $-$17 & 0.511 & 0.4 & 1.2 & \ref{images}.2   \\ 
  & 2007 Mar 28 & BL137P\tablenotemark{a} & 15.4 & 0.86 & 0.61 & $-$15 & 0.602 & 0.3 & 1.0 & \ref{images}.3   \\ 
  & 2007 Aug 24 & BL149AM\tablenotemark{a} & 15.4 & 0.92 & 0.58 & $-$28 & 0.554 & 0.3 & 0.8 & \ref{images}.4   \\ 
  & 2008 May 1 & BL149AO\tablenotemark{a} & 15.4 & 0.82 & 0.57 & $-$9 & 0.806 & 0.2 & 0.7 & \ref{images}.5   \\ 
  & 2008 Jul 17 & BL149AK\tablenotemark{a} & 15.4 & 0.84 & 0.55 & $-$12 & 0.725 & 0.2 & 0.6 & \ref{images}.6   \\ 
  & 2009 Mar 25 & BL149BJ\tablenotemark{a} & 15.4 & 0.84 & 0.62 & $-$12 & 0.435 & 0.2 & 0.5 & \ref{images}.7   \\ 
  & 2010 Jul 12 & BL149CL\tablenotemark{a} & 15.4 & 0.89 & 0.54 & $-$12 & 0.438 & 0.2 & 0.5 & \ref{images}.8   \\ 
0003$-$066  & 2008 Jul 30 & BL149AL\tablenotemark{a} & 15.4 & 1.38 & 0.53 & $-$8 & 1.951 & 0.2 & 0.7 & \ref{images}.9   \\ 
  & 2009 May 2 & BL149BK\tablenotemark{a} & 15.4 & 1.21 & 0.48 & $-$7 & 2.513 & 0.2 & 0.6 & \ref{images}.10   \\ 
  & 2009 Oct 27 & BL149CC\tablenotemark{a} & 15.4 & 1.71 & 0.56 & $-$13 & 2.141 & 0.2 & 1.0 & \ref{images}.11   \\ 
  & 2010 Aug 6 & BL149CM\tablenotemark{a} & 15.4 & 1.33 & 0.59 & 1 & 2.143 & 0.2 & 0.6 & \ref{images}.12   \\ 
  & 2010 Nov 29 & BL149CY\tablenotemark{a} & 15.4 & 1.48 & 0.53 & $-$8 & 2.055 & 0.2 & 0.5 & \ref{images}.13   \\ 
0007+106  & 2008 Aug 25 & BL149BB\tablenotemark{a} & 15.4 & 1.17 & 0.49 & $-$11 & 0.511 & 0.3 & 0.8 & \ref{images}.14 
\enddata 
\tablecomments{Columns are as follows: (1) B1950 name, (2) date of VLBA observation, (3) VLBA experiment code, (4) observing frequency (GHz), (5) FWHM major axis of restoring beam (milliarcseconds), (6) FWHM minor axis of restoring beam (milliarcseconds), (7) position angle of major axis of restoring beam (degrees), (8) total I flux density (Jy),  (9) rms noise level of image (mJy per beam), (10) lowest I contour (mJy per beam), (11) figure number.}

\tablenotetext{a}{Full polarization MOJAVE epoch}
\tablenotetext{b}{2 cm VLBA Survey epoch}

\end{deluxetable*} 

%% file: gaussiantablestub.tex
\begin{deluxetable*}{lclcrrcrcc} 
\tablecolumns{10} 
\tabletypesize{\scriptsize} 
\tablewidth{0pt}  
\tablecaption{\label{3gaussiantable}Fitted Jet Components}  
\tablehead{\colhead{} & \colhead {} &   \colhead {} & 
 \colhead{I} & \colhead{r} &\colhead{P.A.} & \colhead{Maj.} & 
\colhead{} &\colhead{Maj. P.A.}   \\  
\colhead{Source} & \colhead {I.D.} &  \colhead {Epoch} & 
\colhead{(Jy)} & \colhead{(mas)} &\colhead{(\arcdeg)} & \colhead{(mas)} & 
\colhead{Ratio} &\colhead{(\arcdeg)}&\colhead{Robust?}   \\  
\colhead{(1)} & \colhead{(2)} & \colhead{(3)} & \colhead{(4)} &  
\colhead{(5)} & \colhead{(6)} & \colhead{(7)} & \colhead{(8)} & 
 \colhead{(9)} &  \colhead{(10)}} 
\startdata 
0003+380  & 0& 2006 Mar 9  & 0.489  & 0.04 & 290.7 & 0.23 & 0.33 & 292 & Y\\ 
  & 1&   & 0.007  & 3.98 & 121.8 & 0.72 & 1.00 & \n & Y\\ 
  & 2&   & 0.042  & 1.25 & 110.5 & 0.51 & 1.00 & \n & Y\\ 
  & 5&   & 0.104  & 0.28 & 114.6 & 0.27 & 1.00 & \n & Y\\ 
  & 6&   & 0.003  & 2.31 & 119.3 & \n & \n & \n & N\\ 
  & 0& 2006 Dec 1  & 0.320  & 0.10 & 308.1 & 0.25 & 0.29 & 295 & Y\\ 
  & 1&   & 0.005  & 3.65 & 120.8 & 1.63 & 1.00 & \n & Y\\ 
  & 2&   & 0.021  & 1.56 & 111.0 & 0.25 & 1.00 & \n & Y\\ 
  & 4&   & 0.023  & 0.75 & 116.2 & 0.32 & 1.00 & \n & Y\\ 
  & 5&   & 0.145  & 0.45 & 116.3 & 0.05 & 1.00 & \n & Y\\ 
  & 0& 2007 Mar 28  & 0.386  & 0.04 & 309.3 & 0.16 & 0.21 & 307 & Y\\ 
  & 1&   & 0.004  & 4.10 & 119.5 & 0.45 & 1.00 & \n & Y\\ 
  & 2&   & 0.024  & 1.68 & 111.6 & 0.38 & 1.00 & \n & Y\\ 
  & 4&   & 0.053  & 0.72 & 117.1 & 0.15 & 1.00 & \n & Y\\ 
  & 5&   & 0.130  & 0.50 & 115.6 & 0.08 & 1.00 & \n & Y
\enddata 

\tablenotetext{a}{Individual component epoch not used in kinematic fits.}

\tablecomments{Columns are as follows: (1) B1950 name, (2) component identification number (zero indicates core component), (3) observation epoch, (4) flux density in Jy, (5) position offset from the core component (or map center for the core component entries) in milliarcseconds, (6) position angle with respect to the core component (or map center for the core component entries) in degrees,  (7) FWHM major axis of fitted Gaussian in milliarcseconds, (8) axial ratio of fitted Gaussian, (9) major axis position angle of fitted Gaussian in degrees, (10) robust component flag. }

\end{deluxetable*}

%% file: 5velocitytablestub.tex
\begin{deluxetable*}{lcrrrrrrrrrrrrr} 
\tablecolumns{15} 
\tabletypesize{\scriptsize} 
\tablewidth{0pt}  
\tablecaption{\label{5velocitytable}Kinematic Fit Properties of Jet Components}  
\tablehead{\colhead{} & \colhead {} &   \colhead {} & 
\colhead{$\langle S\rangle$}  &\colhead{$\langle R\rangle$} &\colhead{$\langle d_{\mathrm{proj}}\rangle$} & \colhead{$\langle\vartheta\rangle$} & 
 \colhead{$\phi$}&   \colhead{$ |\langle\vartheta\rangle - \phi|$}  &\colhead{$\mu$}  & \colhead{$\beta_{app}$} & & & \colhead{$\Delta \alpha$} &  \colhead{$\Delta \delta$}  \\  
\colhead{Source} & \colhead {I.D.} &  \colhead {N} & 
\colhead{(mJy)} &\colhead{(mas)} & \colhead{(pc)} & \colhead{(deg)}   & 
\colhead{(deg)}& \colhead{(deg)} &\colhead{($\mu$as y$^{-1})$}& \colhead{($c$)}  &\colhead{$T_{ej}$}  & \colhead{$T_{mid}$} & \colhead{($\mu$as)}& \colhead{($\mu$as)}  \\  
\colhead{(1)} & \colhead{(2)} & \colhead{(3)} & \colhead{(4)} &  
\colhead{(5)} & \colhead{(6)} & \colhead{(7)} & \colhead{(8)} & 
 \colhead{(9)}& \colhead{(10)}&  
\colhead{(11)} & \colhead{(12)} & \colhead{(13)} & \colhead{(14)}   & \colhead{(15)} } 
\startdata 
0003+380  & 1 & 8  & 5 &4.2&  15.24& $ 120.8$ &
 99$\pm$19 & 22$\pm$19 & 180$\pm$58 & 2.62$\pm$0.84 & \nodata &2008.35 & 212 &223 \\ 
  & 2 & 7  & 17 &1.8&  6.53& $ 112.6$ &
 119.7$\pm$2.6 & 7.1$\pm$2.7 & 319$\pm$22 & 4.63$\pm$0.32 & 2003.01$\pm$0.24 &2007.70 & 64 &24 \\ 
  & 3 & 5  & 17 &1.3&  4.72& $ 114.2$ &
 202$\pm$15 & 88$\pm$15\tablenotemark{b} & 46$\pm$14 & 0.67$\pm$0.20 & \nodata &2009.08 & 26 &31 \\ 
  & 4 & 7  & 40 &0.8&  2.90& $ 118.2$ &
 254$\pm$28 & 135$\pm$28 & 25$\pm$22 & 0.36$\pm$0.32 & \nodata &2008.72 & 70 &33 \\ 
  & 5\tablenotemark{d} & 8  & 111 &0.4&  1.45& $ 116.6$ &
 336$\pm$102 & 140$\pm$102 & 11$\pm$18 & 0.16$\pm$0.26 & \nodata &2008.35 & 73 &64 \\ 
0003$-$066  & 2 & 5  & 222 &1.0&  4.88& $ 322.9$ &
 226.3$\pm$4.8 & 96.6$\pm$5.0\tablenotemark{b} & 191$\pm$15 & 4.09$\pm$0.33 & \nodata &1997.80 & 19 &81 \\ 
  & 3 & 9  & 119 &2.8&  13.65& $ 296.9$ &
 284.8$\pm$4.5 & 12.1$\pm$4.6 & 250$\pm$39 & 5.35$\pm$0.83 & \nodata &1999.33 & 285 &122 \\ 
  & 4 & 23  & 129 &6.6&  32.19& $ 285.4$ &
 269.7$\pm$2.8 & 15.6$\pm$2.8\tablenotemark{b} & 79.9$\pm$6.2\tablenotemark{a} & 1.71$\pm$0.13 & \nodata &2003.87 & 113 &70 \\ 
  & 5 & 6  & 1031 &0.7&  3.41& $ 14.5$ &
 343.1$\pm$3.1 & 31.4$\pm$3.1\tablenotemark{b} & 100$\pm$16 & 2.15$\pm$0.35 & \nodata &2005.37 & 4 &32 \\ 
  & 6 & 10  & 97 &1.0&  4.88& $ 290.2$ &
 211.3$\pm$8.8 & 78.9$\pm$8.9\tablenotemark{b} & 54$\pm$11\tablenotemark{a} & 1.16$\pm$0.24 & \nodata &2003.78 & 32 &69 \\ 
  & 8 & 9  & 105 &2.2&  10.73& $ 294.6$ &
 292.9$\pm$2.6 & 1.8$\pm$2.7 & 392$\pm$18 & 8.39$\pm$0.39 & 2003.16$\pm$0.27 &2008.96 & 77 &75 \\ 
  & 9 & 9  & 115 &1.6&  7.80& $ 288.2$ &
 300.5$\pm$4.0 & 12.2$\pm$4.1 & 316$\pm$24 & 6.78$\pm$0.52 & \nodata &2008.96 & 105 &86 \\ 
0007+106  & 1 & 8  & 256 &0.5&  0.82& $ 290.7$ &
 291.6$\pm$2.0 & 0.9$\pm$2.4 & 204$\pm$12 & 1.196$\pm$0.069 & 2003.99$\pm$0.16 &2006.77 & 60 &28 \\ 
0010+405  & 1 & 10  & 3 &8.2&  32.35& $ 328.7$ &
 340.8$\pm$4.1 & 12.1$\pm$4.1 & 428$\pm$40 & 6.91$\pm$0.64 & \nodata &2008.52 & 141 &198 \\ 
  & 2\tablenotemark{d} & 11  & 10 &1.7&  6.71& $ 328.6$ &
 354$\pm$111 & 26$\pm$111 & 6$\pm$19 & 0.10$\pm$0.30 & \nodata &2008.62 & 62 &99
\enddata 

\tablenotetext{a}{Component shows significant accelerated motion.}
\tablenotetext{b}{Component shows significant non-radial motion.}
\tablenotetext{c}{Component shows significant inward motion.}
\tablenotetext{d}{Component has a slow pattern speed.}
\tablenotetext{~~}{A question mark indicates a non-radially-moving component for which an inward/outward determination is uncertain due to its slow angular speed.}

\tablecomments{The kinematic fit values are derived from the acceleration fit for components with significant acceleration, and from the vector motion fit otherwise. Columns are as follows: (1) B1950 name, (2) component number, (3) number of fitted epochs, (4) mean flux density at 15 GHz in mJy,  (5) mean distance from core component in mas, (6) mean projected distance from core component in pc, (7) mean position angle with respect to the core component in degrees, (8) position angle of velocity vector in degrees, (9) offset between mean position angle and velocity vector position angle in degrees,  (10) angular proper motion in microarcseconds per year, (11) fitted speed in units of the speed of light, (12) fitted ejection date, (13) date of reference (middle) epoch used for fit, (14) right ascension error of individual epoch positions  in $\mu$as, (15) declination error of individual epoch positions in $\mu$as.  }
\end{deluxetable*}

%% file: accelmotiontablestub.tex
\begin{deluxetable*}{lrrrrrrr} 
\tablecolumns{8} 
\tabletypesize{\scriptsize} 
\tablewidth{0pt}  
\tablecaption{\label{acceltable}Acceleration Fit Properties of Jet Components}  
\tablehead{\colhead{} & \colhead {}& \colhead{$\phi$} & 
\colhead{$\mu$} & \colhead{$ \dot{\mu}_{\perp} $}  & \colhead{$ \dot{\mu}_{\parallel} $}  & &\\  
\colhead{Source} & \colhead {I.D.} & \colhead{(deg)} &   
\colhead{($\mu$as y$^{-1})$} &\colhead{($\mu$as y$^{-2})$}&\colhead{($\mu$as y$^{-2})$}&\colhead{$ \dot{\eta}_{\perp} $}  & \colhead{$ \dot{\eta}_{\parallel} $}  \\  
\colhead{(1)} & \colhead{(2)} & \colhead{(3)} & \colhead{(4)} &  
\colhead{(5)} & \colhead{(6)} & \colhead{(7)} & \colhead{(8)} } 
\startdata 
0003$-$066  & 4\tablenotemark{a}  & $ 269.7 \pm 2.8 $ & $79.9\pm 6.2$ &  $ 9.8 \pm 1.9 $ & $ -24.1 \pm 3.1 $  & 0.165$\pm$0.035 & $-$0.407$\pm$0.061  \\ 
  & 6\tablenotemark{a}  & $ 211.3 \pm 8.8 $ & $54\pm 11$ &  $ 54 \pm 11 $ & $ -37 \pm 15 $  & \n & \n  \\ 
0010+405  & 1  & $ 340.7 \pm 4.3 $ & $428\pm 40$ &  $ 11 \pm 53 $ & $ -43 \pm 69 $  & 0.03$\pm$0.16 & $-$0.13$\pm$0.20  \\ 
  & 2  & $ 354 \pm 132 $ & $6\pm 19$ &  $ 1 \pm 20 $ & $ 4 \pm 33 $  & \n & \n  \\ 
  & 3  & $ 181 \pm 258 $ & $1.6\pm 4.4$ &  $ -8.2 \pm 6.5 $ & $ -2 \pm 10 $  & \n & \n  \\ 
  & 4  & $ 120 \pm 103 $ & $2.4\pm 3.5$ &  $ -2.5 \pm 8.8 $ & $ -3.5 \pm 7.1 $  & \n & \n  \\ 
0016+731  & 1\tablenotemark{a}  & $ 163.2 \pm 2.2 $ & $106.2\pm 4.4$ &  $ 10.2 \pm 1.8 $ & $ 8.8 \pm 1.9 $  & 0.266$\pm$0.048 & 0.231$\pm$0.051  \\ 
0055+300  & 4  & $ 304 \pm 17 $ & $37\pm 11$ &  $ 6.8 \pm 7.3 $ & $ -1.8 \pm 8.0 $  & \n & \n  \\ 
  & 6\tablenotemark{a}  & $ 270 \pm 11 $ & $24.5\pm 3.0$ &  $ 7.6 \pm 3.0 $ & $ 6.7 \pm 2.0 $  & \n & \n  \\ 
  & 10  & $ 321.6 \pm 5.5 $ & $46.9\pm 4.1$ &  $ -0.8 \pm 2.7 $ & $ 0.9 \pm 2.6 $  & \n & \n  \\ 
  & 12  & $ -0 \pm 44 $ & $2.9\pm 2.5$ &  $ 2.7 \pm 1.3 $ & $ -3.7 \pm 1.3 $  & \n & \n  \\ 
  & 13  & $ -0 \pm 39 $ & $3.9\pm 2.6$ &  $ 1.7 \pm 1.7 $ & $ -3.4 \pm 1.2 $  & \n & \n  \\ 
  & 14  & $ -0 \pm 32 $ & $2.9\pm 1.9$ &  $ 1.96 \pm 0.86 $ & $ -2.9 \pm 1.0 $  & \n & \n  \\ 
  & 15  & $ 336 \pm 66 $ & $1.1\pm 1.4$ &  $ -0.47 \pm 0.79 $ & $ -0.55 \pm 0.86 $  & \n & \n  \\ 
0059+581  & 2  & $ 257.4 \pm 3.0 $ & $168\pm 14$ &  $ 8.3 \pm 5.5 $ & $ -16.5 \pm 8.2 $  & 0.081$\pm$0.054 & $-$0.161$\pm$0.081
\enddata 
\tablenotetext{a}{Component shows significant accelerated motion.}
\tablecomments{Columns are as follows: (1) B1950 name, (2) component number, (3) proper motion position angle in degrees, (4) angular proper motion in microarcseconds per year, (5) angular acceleration perpendicular to velocity direction in microarcseconds per year per year, (6) angular acceleration parallel to velocity direction in microarcseconds per year per year, (7) relative acceleration in direction perpendicular to fitted velocity vector, (8) relative acceleration in direction parallel to fitted velocity vector.}

\end{deluxetable*}

%% file: MOJAVE_paperX.bbl
\begin{thebibliography}{89}
\expandafter\ifx\csname natexlab\endcsname\relax\def\natexlab#1{#1}\fi

\bibitem[{{Abdo} {et~al.}(2010){Abdo}, {Ackermann}, {Ajello}, {Allafort},
  {Antolini}, {Atwood}, {Axelsson}, {Baldini}, {Ballet}, {Barbiellini},
  {Bastieri}, {Baughman}, {Bechtol}, {Bellazzini}, {Berenji}, {Blandford},
  {Bloom}, {Bogart}, {Bonamente}, {Borgland}, {Bouvier}, {Bregeon}, {Brez},
  {Brigida}, {Bruel}, {Buehler}, {Burnett}, {Buson}, {Caliandro}, {Cameron},
  {Cannon}, {Caraveo}, {Carrigan}, {Casandjian}, {Cavazzuti}, {Cecchi}, {{\c
  C}elik}, {Celotti}, {Charles}, {Chekhtman}, {Chen}, {Cheung}, {Chiang},
  {Ciprini}, {Claus}, {Cohen-Tanugi}, {Conrad}, {Costamante}, {Cotter},
  {Cutini}, {D'Elia}, {Dermer}, {de Angelis}, {de Palma}, {De Rosa}, {Digel},
  {Silva}, {Drell}, {Dubois}, {Dumora}, {Escande}, {Farnier}, {Favuzzi},
  {Fegan}, {Ferrara}, {Focke}, {Fortin}, {Frailis}, {Fukazawa}, {Funk},
  {Fusco}, {Gargano}, {Gasparrini}, {Gehrels}, {Germani}, {Giebels},
  {Giglietto}, {Giommi}, {Giordano}, {Giroletti}, {Glanzman}, {Godfrey},
  {Grandi}, {Grenier}, {Grondin}, {Grove}, {Guiriec}, {Hadasch}, {Harding},
  {Hayashida}, {Hays}, {Healey}, {Hill}, {Horan}, {Hughes}, {Iafrate}, {Itoh},
  {J{\'o}hannesson}, {Johnson}, {Johnson}, {Johnson}, {Johnson}, {Kamae},
  {Katagiri}, {Kataoka}, {Kawai}, {Kerr}, {Kn{\"o}dlseder}, {Kuss}, {Lande},
  {Latronico}, {Lavalley}, {Lemoine-Goumard}, {Llena Garde}, {Longo},
  {Loparco}, {Lott}, {Lovellette}, {Lubrano}, {Madejski}, {Makeev}, {Malaguti},
  {Massaro}, {Mazziotta}, {McConville}, {McEnery}, {McGlynn}, {Michelson},
  {Mitthumsiri}, {Mizuno}, {Moiseev}, {Monte}, {Monzani}, {Morselli},
  {Moskalenko}, {Murgia}, {Nolan}, {Norris}, {Nuss}, {Ohno}, {Ohsugi},
  {Omodei}, {Orlando}, {Ormes}, {Ozaki}, {Paneque}, {Panetta}, {Parent},
  {Pelassa}, {Pepe}, {Pesce-Rollins}, {Piranomonte}, {Piron}, {Porter},
  {Rain{\`o}}, {Rando}, {Razzano}, {Reimer}, {Reimer}, {Reposeur}, {Ripken},
  {Ritz}, {Rodriguez}, {Romani}, {Roth}, {Ryde}, {Sadrozinski}, {Sanchez},
  {Sander}, {Saz Parkinson}, {Scargle}, {Sgr{\`o}}, {Shaw}, {Siskind}, {Smith},
  {Spandre}, {Spinelli}, {Starck}, {Stawarz}, {Strickman}, {Suson}, {Tajima},
  {Takahashi}, {Takahashi}, {Tanaka}, {Taylor}, {Thayer}, {Thayer}, {Thompson},
  {Tibaldo}, {Torres}, {Tosti}, {Tramacere}, {Ubertini}, {Uchiyama}, {Usher},
  {Vasileiou}, {Vilchez}, {Villata}, {Vitale}, {Waite}, {Wallace}, {Wang},
  {Winer}, {Wood}, {Yang}, {Ylinen}, \& {Ziegler}}]{1LAC}
{Abdo}, A.~A., {et~al.} 2010, \apj, 715, 429

\bibitem[{{Ackermann} {et~al.}(2011){Ackermann}, {Ajello}, {Allafort},
  {Antolini}, {Atwood}, {Axelsson}, {Baldini}, {Ballet}, {Barbiellini},
  {Bastieri}, {Bechtol}, {Bellazzini}, {Berenji}, {Blandford}, {Bloom},
  {Bonamente}, {Borgland}, {Bottacini}, {Bouvier}, {Bregeon}, {Brigida},
  {Bruel}, {Buehler}, {Burnett}, {Buson}, {Caliandro}, {Cameron}, {Caraveo},
  {Casandjian}, {Cavazzuti}, {Cecchi}, {Charles}, {Cheung}, {Chiang},
  {Ciprini}, {Claus}, {Cohen-Tanugi}, {Conrad}, {Costamante}, {Cutini}, {de
  Angelis}, {de Palma}, {Dermer}, {Digel}, {Silva}, {Drell}, {Dubois},
  {Escande}, {Favuzzi}, {Fegan}, {Ferrara}, {Finke}, {Focke}, {Fortin},
  {Frailis}, {Fukazawa}, {Funk}, {Fusco}, {Gargano}, {Gasparrini}, {Gehrels},
  {Germani}, {Giebels}, {Giglietto}, {Giommi}, {Giordano}, {Giroletti},
  {Glanzman}, {Godfrey}, {Grenier}, {Grove}, {Guiriec}, {Gustafsson},
  {Hadasch}, {Hayashida}, {Hays}, {Healey}, {Horan}, {Hou}, {Hughes},
  {Iafrate}, {J{\'o}hannesson}, {Johnson}, {Johnson}, {Kamae}, {Katagiri},
  {Kataoka}, {Kn{\"o}dlseder}, {Kuss}, {Lande}, {Larsson}, {Latronico},
  {Longo}, {Loparco}, {Lott}, {Lovellette}, {Lubrano}, {Madejski}, {Mazziotta},
  {McConville}, {McEnery}, {Michelson}, {Mitthumsiri}, {Mizuno}, {Moiseev},
  {Monte}, {Monzani}, {Moretti}, {Morselli}, {Moskalenko}, {Murgia},
  {Nakamori}, {Naumann-Godo}, {Nolan}, {Norris}, {Nuss}, {Ohno}, {Ohsugi},
  {Okumura}, {Omodei}, {Orienti}, {Orlando}, {Ormes}, {Ozaki}, {Paneque},
  {Parent}, {Pesce-Rollins}, {Pierbattista}, {Piranomonte}, {Piron}, {Pivato},
  {Porter}, {Rain{\`o}}, {Rando}, {Razzano}, {Razzaque}, {Reimer}, {Reimer},
  {Ritz}, {Rochester}, {Romani}, {Roth}, {Sanchez}, {Sbarra}, {Scargle},
  {Schalk}, {Sgr{\`o}}, {Shaw}, {Siskind}, {Spandre}, {Spinelli}, {Strong},
  {Suson}, {Tajima}, {Takahashi}, {Takahashi}, {Tanaka}, {Thayer}, {Thayer},
  {Thompson}, {Tibaldo}, {Tinivella}, {Torres}, {Tosti}, {Troja}, {Uchiyama},
  {Vandenbroucke}, {Vasileiou}, {Vianello}, {Vitale}, {Waite}, {Wallace},
  {Wang}, {Winer}, {Wood}, {Wood}, \& {Zimmer}}]{2LAC}
{Ackermann}, M., {et~al.} 2011, \apj, 743, 171

\bibitem[{{Agudo}(2009)}]{2009ASPC..402..330A}
{Agudo}, I. 2009, in Astronomical Society of the Pacific Conference Series,
  Vol. 402, Approaching Micro-Arcsecond Resolution with VSOP-2: Astrophysics
  and Technologies, ed. Y.~{Hagiwara}, E.~{Fomalont}, M.~{Tsuboi}, \&
  M.~{Yasuhiro}, 330

\bibitem[{{Agudo} {et~al.}(2012){Agudo}, {Marscher}, {Jorstad}, {G{\'o}mez},
  {Perucho}, {Piner}, {Rioja}, \& {Dodson}}]{2012ApJ...747...63A}
{Agudo}, I., {Marscher}, A.~P., {Jorstad}, S.~G., {G{\'o}mez}, J.~L.,
  {Perucho}, M., {Piner}, B.~G., {Rioja}, M., \& {Dodson}, R. 2012, \apj, 747,
  63

\bibitem[{{Agudo} {et~al.}(2007){Agudo}, {Bach}, {Krichbaum}, {Marscher},
  {Gonidakis}, {Diamond}, {Perucho}, {Alef}, {Graham}, {Witzel}, {Zensus},
  {Bremer}, {Acosta-Pulido}, \& {Barrena}}]{2007AA...476L..17A}
{Agudo}, I., {et~al.} 2007, \aap, 476, L17

\bibitem[{{Albert} {et~al.}(2007){Albert}, {Aliu}, {Anderhub}, {Antoranz},
  {Armada}, {Baixeras}, {Barrio}, {Bartko}, {Bastieri}, {Becker}, {Bednarek},
  {Berger}, {Bigongiari}, {Biland}, {Bock}, {Bordas}, {Bosch-Ramon}, {Bretz},
  {Britvitch}, {Camara}, {Carmona}, {Chilingarian}, {Coarasa}, {Commichau},
  {Contreras}, {Cortina}, {Costado}, {Curtef}, {Danielyan}, {Dazzi}, {De
  Angelis}, {Delgado}, {de los Reyes}, {De Lotto}, {Domingo-Santamar{\'{\i}}a},
  {Dorner}, {Doro}, {Errando}, {Fagiolini}, {Ferenc}, {Fern{\'a}ndez}, {Firpo},
  {Flix}, {Fonseca}, {Font}, {Fuchs}, {Galante}, {Garc{\'{\i}}a-L{\'o}pez},
  {Garczarczyk}, {Gaug}, {Giller}, {Goebel}, {Hakobyan}, {Hayashida},
  {Hengstebeck}, {Herrero}, {H{\"o}hne}, {Hose}, {Hsu}, {Jacon}, {Jogler},
  {Kosyra}, {Kranich}, {Kritzer}, {Laille}, {Lindfors}, {Lombardi}, {Longo},
  {L{\'o}pez}, {L{\'o}pez}, {Lorenz}, {Majumdar}, {Maneva}, {Mannheim},
  {Mansutti}, {Mariotti}, {Mart{\'{\i}}nez}, {Mazin}, {Merck}, {Meucci},
  {Meyer}, {Miranda}, {Mirzoyan}, {Mizobuchi}, {Moralejo}, {Nieto}, {Nilsson},
  {Ninkovic}, {O{\~n}a-Wilhelmi}, {Otte}, {Oya}, {Paneque}, {Panniello},
  {Paoletti}, {Paredes}, {Pasanen}, {Pascoli}, {Pauss}, {Pegna}, {Perlman},
  {Persic}, {Peruzzo}, {Piccioli}, {Prandini}, {Puchades}, {Raymers}, {Rhode},
  {Rib{\'o}}, {Rico}, {Rissi}, {Robert}, {R{\"u}gamer}, {Saggion}, {Saito},
  {S{\'a}nchez}, {Sartori}, {Scalzotto}, {Scapin}, {Schmitt}, {Schweizer},
  {Shayduk}, {Shinozaki}, {Shore}, {Sidro}, {Sillanp{\"a}{\"a}}, {Sobczynska},
  {Stamerra}, {Stark}, {Takalo}, {Tavecchio}, {Temnikov}, {Tescaro}, {Teshima},
  {Torres}, {Turini}, {Vankov}, {Vitale}, {Wagner}, {Wibig}, {Wittek},
  {Zandanel}, {Zanin}, \& {Zapatero}}]{2007ApJ...667L..21A}
{Albert}, J., {et~al.} 2007, \apjl, 667, L21

\bibitem[{{Asada} {et~al.}(2008){Asada}, {Inoue}, {Kameno}, \&
  {Nagai}}]{2008ApJ...675...79A}
{Asada}, K., {Inoue}, M., {Kameno}, S., \& {Nagai}, H. 2008, \apj, 675, 79

\bibitem[{{Bach} {et~al.}(2005){Bach}, {Krichbaum}, {Ros}, {Britzen}, {Tian},
  {Kraus}, {Witzel}, \& {Zensus}}]{2005AA...433..815B}
{Bach}, U., {Krichbaum}, T.~P., {Ros}, E., {Britzen}, S., {Tian}, W.~W.,
  {Kraus}, A., {Witzel}, A., \& {Zensus}, J.~A. 2005, \aap, 433, 815

\bibitem[{{Beasley} {et~al.}(2002){Beasley}, {Gordon}, {Peck}, {Petrov},
  {MacMillan}, {Fomalont}, \& {Ma}}]{2002ApJS..141...13B}
{Beasley}, A.~J., {Gordon}, D., {Peck}, A.~B., {Petrov}, L., {MacMillan},
  D.~S., {Fomalont}, E.~B., \& {Ma}, C. 2002, \apjs, 141, 13

\bibitem[{{Cara} \& {Lister}(2008)}]{MOJAVE_IV}
{Cara}, M., \& {Lister}, M.~L. 2008, \apj, 674, 111

\bibitem[{{Chang} {et~al.}(2010){Chang}, {Ros}, {Kovalev}, \&
  {Lister}}]{2010AA...515A..38C}
{Chang}, C.~S., {Ros}, E., {Kovalev}, Y.~Y., \& {Lister}, M.~L. 2010, \aap,
  515, A38

\bibitem[{{Cohen} {et~al.}(2007){Cohen}, {Lister}, {Homan}, {Kadler},
  {Kellermann}, {Kovalev}, \& {Vermeulen}}]{2007ApJ...658..232C}
{Cohen}, M.~H., {Lister}, M.~L., {Homan}, D.~C., {Kadler}, M., {Kellermann},
  K.~I., {Kovalev}, Y.~Y., \& {Vermeulen}, R.~C. 2007, \apj, 658, 232

\bibitem[{{Danforth} {et~al.}(2013){Danforth}, {Nalewajko}, {France}, \&
  {Keeney}}]{2013ApJ...764...57D}
{Danforth}, C.~W., {Nalewajko}, K., {France}, K., \& {Keeney}, B.~A. 2013,
  \apj, 764, 57

\bibitem[{{Deller} {et~al.}(2011){Deller}, {Brisken}, {Phillips}, {Morgan},
  {Alef}, {Cappallo}, {Middelberg}, {Romney}, {Rottmann}, {Tingay}, \&
  {Wayth}}]{2011PASP..123..275D}
{Deller}, A.~T., {et~al.} 2011, \pasp, 123, 275

\bibitem[{{Duncan} \& {Hughes}(1994)}]{1994ApJ...436L.119D}
{Duncan}, G.~C., \& {Hughes}, P.~A. 1994, \apjl, 436, L119

\bibitem[{{Ellerbroek} {et~al.}(2013){Ellerbroek}, {Podio}, {Kaper}, {Sana},
  {Huppenkothen}, {de Koter}, \& {Monaco}}]{2013AA...551A...5E}
{Ellerbroek}, L.~E., {Podio}, L., {Kaper}, L., {Sana}, H., {Huppenkothen}, D.,
  {de Koter}, A., \& {Monaco}, L. 2013, \aap, 551, A5

\bibitem[{{Fomalont}(1999)}]{1999ASPC..180..301F}
{Fomalont}, E.~B. 1999, in Astronomical Society of the Pacific Conference
  Series, Vol. 180, Synthesis Imaging in Radio Astronomy II, ed. G.~B.
  {Taylor}, C.~L. {Carilli}, \& R.~A. {Perley}, 301

\bibitem[{{Fomalont} {et~al.}(2003){Fomalont}, {Petrov}, {MacMillan}, {Gordon},
  \& {Ma}}]{2003AJ....126.2562F}
{Fomalont}, E.~B., {Petrov}, L., {MacMillan}, D.~S., {Gordon}, D., \& {Ma}, C.
  2003, \aj, 126, 2562

\bibitem[{{Furniss} {et~al.}(2013){Furniss}, {Fumagalli}, {Danforth},
  {Williams}, \& {Prochaska}}]{2013ApJ...766...35F}
{Furniss}, A., {Fumagalli}, M., {Danforth}, C., {Williams}, D.~A., \&
  {Prochaska}, J.~X. 2013, \apj, 766, 35

\bibitem[{{Giommi} {et~al.}(2012){Giommi}, {Padovani}, {Polenta}, {Turriziani},
  {D'Elia}, \& {Piranomonte}}]{2012MNRAS.420.2899G}
{Giommi}, P., {Padovani}, P., {Polenta}, G., {Turriziani}, S., {D'Elia}, V., \&
  {Piranomonte}, S. 2012, \mnras, 420, 2899

\bibitem[{{Giovannini} {et~al.}(2001){Giovannini}, {Cotton}, {Feretti}, {Lara},
  \& {Venturi}}]{2001ApJ...552..508G}
{Giovannini}, G., {Cotton}, W.~D., {Feretti}, L., {Lara}, L., \& {Venturi}, T.
  2001, \apj, 552, 508

\bibitem[{{G{\'o}mez} {et~al.}(2011){G{\'o}mez}, {Roca-Sogorb}, {Agudo},
  {Marscher}, \& {Jorstad}}]{2011ApJ...733...11G}
{G{\'o}mez}, J.~L., {Roca-Sogorb}, M., {Agudo}, I., {Marscher}, A.~P., \&
  {Jorstad}, S.~G. 2011, \apj, 733, 11

\bibitem[{{Hardee}(2011)}]{2011IAUS..275...41H}
{Hardee}, P.~E. 2011, in IAU Symposium, Vol. 275, IAU Symposium, ed. G.~E.
  {Romero}, R.~A. {Sunyaev}, \& T.~{Belloni}, 41--49

\bibitem[{{Hewett} {et~al.}(1995){Hewett}, {Foltz}, \&
  {Chaffee}}]{1995AJ....109.1498H}
{Hewett}, P.~C., {Foltz}, C.~B., \& {Chaffee}, F.~H. 1995, \aj, 109, 1498

\bibitem[{{Homan} {et~al.}(2009){Homan}, {Kadler}, {Kellermann}, {Kovalev},
  {Lister}, {Ros}, {Savolainen}, \& {Zensus}}]{MOJAVE_VII}
{Homan}, D.~C., {Kadler}, M., {Kellermann}, K.~I., {Kovalev}, Y.~Y., {Lister},
  M.~L., {Ros}, E., {Savolainen}, T., \& {Zensus}, J.~A. 2009, \apj, 706, 1253

\bibitem[{{Homan} {et~al.}(2002){Homan}, {Ojha}, {Wardle}, {Roberts}, {Aller},
  {Aller}, \& {Hughes}}]{2002ApJ...568...99H}
{Homan}, D.~C., {Ojha}, R., {Wardle}, J.~F.~C., {Roberts}, D.~H., {Aller},
  M.~F., {Aller}, H.~D., \& {Hughes}, P.~A. 2002, \apj, 568, 99

\bibitem[{{Hook} {et~al.}(1996){Hook}, {McMahon}, {Irwin}, \&
  {Hazard}}]{1996MNRAS.282.1274H}
{Hook}, I.~M., {McMahon}, R.~G., {Irwin}, M.~J., \& {Hazard}, C. 1996, \mnras,
  282, 1274

\bibitem[{{Hovatta} {et~al.}(2012){Hovatta}, {Lister}, {Aller}, {Aller},
  {Homan}, {Kovalev}, {Pushkarev}, \& {Savolainen}}]{2012AJ....144..105H}
{Hovatta}, T., {Lister}, M.~L., {Aller}, M.~F., {Aller}, H.~D., {Homan}, D.~C.,
  {Kovalev}, Y.~Y., {Pushkarev}, A.~B., \& {Savolainen}, T. 2012, \aj, 144, 105

\bibitem[{{Huchra} {et~al.}(1999){Huchra}, {Vogeley}, \&
  {Geller}}]{1999ApJS..121..287H}
{Huchra}, J.~P., {Vogeley}, M.~S., \& {Geller}, M.~J. 1999, \apjs, 121, 287

\bibitem[{{Jones} {et~al.}(2005){Jones}, {Saunders}, {Read}, \&
  {Colless}}]{2005PASA...22..277J}
{Jones}, D.~H., {Saunders}, W., {Read}, M., \& {Colless}, M. 2005, PASA, 22,
  277

\bibitem[{{Jorstad} {et~al.}(2004){Jorstad}, {Marscher}, {Lister}, {Stirling},
  {Cawthorne}, {G{\'o}mez}, \& {Gear}}]{2004AJ....127.3115J}
{Jorstad}, S.~G., {Marscher}, A.~P., {Lister}, M.~L., {Stirling}, A.~M.,
  {Cawthorne}, T.~V., {G{\'o}mez}, J.-L., \& {Gear}, W.~K. 2004, \aj, 127, 3115

\bibitem[{{Jorstad} {et~al.}(2005){Jorstad}, {Marscher}, {Lister}, {Stirling},
  {Cawthorne}, {Gear}, {G{\'o}mez}, {Stevens}, {Smith}, {Forster}, \&
  {Robson}}]{2005AJ....130.1418J}
{Jorstad}, S.~G., {et~al.} 2005, \aj, 130, 1418

\bibitem[{{Kadler} {et~al.}(2004){Kadler}, {Ros}, {Lobanov}, {Falcke}, \&
  {Zensus}}]{2004AA...426..481K}
{Kadler}, M., {Ros}, E., {Lobanov}, A.~P., {Falcke}, H., \& {Zensus}, J.~A.
  2004, \aap, 426, 481

\bibitem[{{Kellermann} {et~al.}(1998){Kellermann}, {Vermeulen}, {Zensus}, \&
  {Cohen}}]{2cmPaperI}
{Kellermann}, K.~I., {Vermeulen}, R.~C., {Zensus}, J.~A., \& {Cohen}, M.~H.
  1998, \aj, 115, 1295

\bibitem[{{Kellermann} {et~al.}(2004){Kellermann}, {Lister}, {Homan},
  {Vermeulen}, {Cohen}, {Ros}, {Kadler}, {Zensus}, \&
  {Kovalev}}]{2004ApJ...609..539K}
{Kellermann}, K.~I., {et~al.} 2004, \apj, 609, 539

\bibitem[{{Kovalev} {et~al.}(2007){Kovalev}, {Petrov}, {Fomalont}, \&
  {Gordon}}]{2007AJ....133.1236K}
{Kovalev}, Y.~Y., {Petrov}, L., {Fomalont}, E.~B., \& {Gordon}, D. 2007, \aj,
  133, 1236

\bibitem[{{Landoni} {et~al.}(2012){Landoni}, {Falomo}, {Treves}, {Sbarufatti},
  {Decarli}, {Tavecchio}, \& {Kotilainen}}]{2012AA...543A.116L}
{Landoni}, M., {Falomo}, R., {Treves}, A., {Sbarufatti}, B., {Decarli}, R.,
  {Tavecchio}, F., \& {Kotilainen}, J. 2012, \aap, 543, A116

\bibitem[{{Larionov} {et~al.}(2013){Larionov}, {Jorstad}, {Marscher},
  {Morozova}, {Blinov}, {Hagen-Thorn}, {Konstantinova}, {Kopatskaya},
  {Larionova}, {Larionova}, \& {Troitsky}}]{2013ApJ...768...40L}
{Larionov}, V.~M., {et~al.} 2013, \apj, 768, 40

\bibitem[{{Lawrence} {et~al.}(1986){Lawrence}, {Pearson}, {Readhead}, \&
  {Unwin}}]{1986AJ.....91..494L}
{Lawrence}, C.~R., {Pearson}, T.~J., {Readhead}, A.~C.~S., \& {Unwin}, S.~C.
  1986, \aj, 91, 494

\bibitem[{{Lister} \& {Homan}(2005)}]{MOJAVE_I}
{Lister}, M.~L., \& {Homan}, D.~C. 2005, \aj, 130, 1389

\bibitem[{{Lister} \& {Marscher}(1997)}]{LM97}
{Lister}, M.~L., \& {Marscher}, A.~P. 1997, \apj, 476, 572

\bibitem[{{Lister} {et~al.}(2009{\natexlab{a}}){Lister}, {Aller}, {Aller},
  {Cohen}, {Homan}, {Kadler}, {Kellermann}, {Kovalev}, {Ros}, {Savolainen},
  {Zensus}, \& {Vermeulen}}]{MOJAVE_V}
{Lister}, M.~L., {et~al.} 2009{\natexlab{a}}, \aj, 137, 3718

\bibitem[{{Lister} {et~al.}(2009{\natexlab{b}}){Lister}, {Cohen}, {Homan},
  {Kadler}, {Kellermann}, {Kovalev}, {Ros}, {Savolainen}, \&
  {Zensus}}]{MOJAVE_VI}
---. 2009{\natexlab{b}}, \aj, 138, 1874

\bibitem[{{Lister} {et~al.}(2011){Lister}, {Aller}, {Aller}, {Hovatta},
  {Kellermann}, {Kovalev}, {Meyer}, {Pushkarev}, {Ros}, {MOJAVE Collaboration},
  {Ackermann}, {Antolini}, {Baldini}, {Ballet}, {Barbiellini}, {Bastieri},
  {Bechtol}, {Bellazzini}, {Berenji}, {Blandford}, {Bloom}, {Boeck},
  {Bonamente}, {Borgland}, {Bregeon}, {Brigida}, {Bruel}, {Buehler}, {Buson},
  {Caliandro}, {Cameron}, {Caraveo}, {Casandjian}, {Cavazzuti}, {Cecchi},
  {Chang}, {Charles}, {Chekhtman}, {Cheung}, {Chiang}, {Ciprini}, {Claus},
  {Cohen-Tanugi}, {Conrad}, {Cutini}, {de Palma}, {Dermer}, {Silva}, {Drell},
  {Drlica-Wagner}, {Favuzzi}, {Fegan}, {Ferrara}, {Finke}, {Focke}, {Fortin},
  {Fukazawa}, {Fusco}, {Gargano}, {Gasparrini}, {Gehrels}, {Germani},
  {Giglietto}, {Giordano}, {Giroletti}, {Glanzman}, {Godfrey}, {Grenier},
  {Guiriec}, {Hadasch}, {Hayashida}, {Hays}, {Horan}, {Hughes},
  {J{\'o}hannesson}, {Johnson}, {Kadler}, {Katagiri}, {Kataoka},
  {Kn{\"o}dlseder}, {Kuss}, {Lande}, {Longo}, {Loparco}, {Lott}, {Lovellette},
  {Lubrano}, {Madejski}, {Mazziotta}, {McConville}, {McEnery}, {Mehault},
  {Michelson}, {Mizuno}, {Monte}, {Monzani}, {Morselli}, {Moskalenko},
  {Murgia}, {Naumann-Godo}, {Nishino}, {Nolan}, {Norris}, {Nuss}, {Ohno},
  {Ohsugi}, {Okumura}, {Omodei}, {Orlando}, {Ozaki}, {Paneque}, {Parent},
  {Pesce-Rollins}, {Pierbattista}, {Piron}, {Pivato}, {Rain{\`o}}, {Readhead},
  {Reimer}, {Reimer}, {Richards}, {Ritz}, {Sadrozinski}, {Sgr{\`o}}, {Shaw},
  {Siskind}, {Spandre}, {Spinelli}, {Takahashi}, {Tanaka}, {Thayer}, {Thayer},
  {Thompson}, {Tosti}, {Tramacere}, {Troja}, {Usher}, {Vandenbroucke},
  {Vasileiou}, {Vianello}, {Vitale}, {Waite}, {Wang}, {Winer}, {Wood},
  {Zimmer}, \& {Fermi LAT Collaboration}}]{2011ApJ...742...27L}
---. 2011, \apj, 742, 27

\bibitem[{{Lobanov} \& {Roland}(2005)}]{2005AA...431..831L}
{Lobanov}, A.~P., \& {Roland}, J. 2005, \aap, 431, 831

\bibitem[{{Lobanov} \& {Zensus}(2001)}]{2001Sci...294..128L}
{Lobanov}, A.~P., \& {Zensus}, J.~A. 2001, Science, 294, 128

\bibitem[{{Lomb}(1976)}]{1976ApSS..39..447L}
{Lomb}, N.~R. 1976, \apss, 39, 447

\bibitem[{{Ly} {et~al.}(2007){Ly}, {Walker}, \& {Junor}}]{2007ApJ...660..200L}
{Ly}, C., {Walker}, R.~C., \& {Junor}, W. 2007, \apj, 660, 200

\bibitem[{{Marti-Vidal} {et~al.}(2013){Marti-Vidal}, {Marcaide}, {Alberdi}, \&
  {Brunthaler}}]{2013arXiv1301.4782M}
{Marti-Vidal}, I., {Marcaide}, J.~M., {Alberdi}, A., \& {Brunthaler}, A. 2013,
  ArXiv e-prints

\bibitem[{{McConville} {et~al.}(2011){McConville}, {Ostorero}, {Moderski},
  {Stawarz}, {Cheung}, {Ajello}, {Bouvier}, {Bregeon}, {Donato}, {Finke},
  {Furniss}, {McEnery}, {Monzani}, {Orienti}, {Reyes}, {Rossetti}, \&
  {Williams}}]{2011ApJ...738..148M}
{McConville}, W., {et~al.} 2011, \apj, 738, 148

\bibitem[{{Meier} {et~al.}(2001){Meier}, {Koide}, \&
  {Uchida}}]{2001Sci...291...84M}
{Meier}, D.~L., {Koide}, S., \& {Uchida}, Y. 2001, Science, 291, 84

\bibitem[{{Meisner} \& {Romani}(2010)}]{2010ApJ...712...14M}
{Meisner}, A.~M., \& {Romani}, R.~W. 2010, \apj, 712, 14

\bibitem[{{Murphy} {et~al.}(1993){Murphy}, {Browne}, \&
  {Perley}}]{1993MNRAS.264..298M}
{Murphy}, D.~W., {Browne}, I.~W.~A., \& {Perley}, R.~A. 1993, \mnras, 264, 298

\bibitem[{{Natarajan} \& {Pringle}(1998)}]{1998ApJ...506L..97N}
{Natarajan}, P., \& {Pringle}, J.~E. 1998, \apjl, 506, L97

\bibitem[{{Nilsson} {et~al.}(2008){Nilsson}, {Pursimo}, {Sillanp{\"a}{\"a}},
  {Takalo}, \& {Lindfors}}]{2008AA...487L..29N}
{Nilsson}, K., {Pursimo}, T., {Sillanp{\"a}{\"a}}, A., {Takalo}, L.~O., \&
  {Lindfors}, E. 2008, \aap, 487, L29

\bibitem[{{Orr} \& {Browne}(1982)}]{1982MNRAS.200.1067O}
{Orr}, M.~J.~L., \& {Browne}, I.~W.~A. 1982, \mnras, 200, 1067

\bibitem[{{Perucho} {et~al.}(2012){Perucho}, {Kovalev}, {Lobanov}, {Hardee}, \&
  {Agudo}}]{2012ApJ...749...55P}
{Perucho}, M., {Kovalev}, Y.~Y., {Lobanov}, A.~P., {Hardee}, P.~E., \& {Agudo},
  I. 2012, \apj, 749, 55

\bibitem[{{Petrov} {et~al.}(2005){Petrov}, {Kovalev}, {Fomalont}, \&
  {Gordon}}]{2005AJ....129.1163P}
{Petrov}, L., {Kovalev}, Y.~Y., {Fomalont}, E., \& {Gordon}, D. 2005, \aj, 129,
  1163

\bibitem[{{Petrov} {et~al.}(2006){Petrov}, {Kovalev}, {Fomalont}, \&
  {Gordon}}]{2006AJ....131.1872P}
{Petrov}, L., {Kovalev}, Y.~Y., {Fomalont}, E.~B., \& {Gordon}, D. 2006, \aj,
  131, 1872

\bibitem[{{Petrov} {et~al.}(2008){Petrov}, {Kovalev}, {Fomalont}, \&
  {Gordon}}]{VCS6}
---. 2008, \aj, 136, 580

\bibitem[{{Piner} {et~al.}(2010){Piner}, {Pant}, \&
  {Edwards}}]{2010ApJ...723.1150P}
{Piner}, B.~G., {Pant}, N., \& {Edwards}, P.~G. 2010, \apj, 723, 1150

\bibitem[{{Pushkarev} {et~al.}(2012){Pushkarev}, {Hovatta}, {Kovalev},
  {Lister}, {Lobanov}, {Savolainen}, \& {Zensus}}]{2012AA...545A.113P}
{Pushkarev}, A.~B., {Hovatta}, T., {Kovalev}, Y.~Y., {Lister}, M.~L.,
  {Lobanov}, A.~P., {Savolainen}, T., \& {Zensus}, J.~A. 2012, \aap, 545, A113

\bibitem[{{Pushkarev} {et~al.}(2009){Pushkarev}, {Kovalev}, {Lister}, \&
  {Savolainen}}]{2009AA...507L..33P}
{Pushkarev}, A.~B., {Kovalev}, Y.~Y., {Lister}, M.~L., \& {Savolainen}, T.
  2009, \aap, 507, L33

\bibitem[{{Pushkarev} {et~al.}(2013){Pushkarev}, {Kovalev}, {Lister},
  {Hovatta}, {Savolainen}, {Aller}, {Aller}, {Ros}, {Zensus}, {Richards},
  {Max-Moerbeck}, \& {Readhead}}]{2013arXiv1305.6005P}
{Pushkarev}, A.~B., {et~al.} 2013, ArXiv e-prints 1305.6005

\bibitem[{{Rani} {et~al.}(2013){Rani}, {Krichbaum}, {Fuhrmann}, {B{\"o}ttcher},
  {Lott}, {Aller}, {Aller}, {Angelakis}, {Bach}, {Bastieri}, {Falcone},
  {Fukazawa}, {Gabanyi}, {Gupta}, {Gurwell}, {Itoh}, {Kawabata}, {Krips},
  {L{\"a}hteenm{\"a}ki}, {Liu}, {Marchili}, {Max-Moerbeck}, {Nestoras},
  {Nieppola}, {Quintana-Lacaci}, {Readhead}, {Richards}, {Sasada}, {Sievers},
  {Sokolovsky}, {Stroh}, {Tammi}, {Tornikoski}, {Uemura}, {Ungerechts},
  {Urano}, \& {Zensus}}]{2013AA...552A..11R}
{Rani}, B., {et~al.} 2013, \aap, 552, A11

\bibitem[{{Rau} {et~al.}(2012){Rau}, {Schady}, {Greiner}, {Salvato}, {Ajello},
  {Bottacini}, {Gehrels}, {Afonso}, {Elliott}, {Filgas}, {Kann}, {Klose},
  {Kr{\"u}hler}, {Nardini}, {Nicuesa Guelbenzu}, {Olivares E.}, {Rossi},
  {Sudilovsky}, {Updike}, \& {Hartmann}}]{2012AA...538A..26R}
{Rau}, A., {et~al.} 2012, \aap, 538, A26

\bibitem[{{Rosen} {et~al.}(1999){Rosen}, {Hughes}, {Duncan}, \&
  {Hardee}}]{1999ApJ...516..729R}
{Rosen}, A., {Hughes}, P.~A., {Duncan}, G.~C., \& {Hardee}, P.~E. 1999, \apj,
  516, 729

\bibitem[{{Sargent}(1970)}]{1970ApJ...160..405S}
{Sargent}, W.~L.~W. 1970, \apj, 160, 405

\bibitem[{{Savolainen} {et~al.}(2006){Savolainen}, {Wiik}, {Valtaoja}, \&
  {Tornikoski}}]{2006AA...446...71S}
{Savolainen}, T., {Wiik}, K., {Valtaoja}, E., \& {Tornikoski}, M. 2006, \aap,
  446, 71

\bibitem[{{Sbarufatti} {et~al.}(2005){Sbarufatti}, {Treves}, \&
  {Falomo}}]{2005ApJ...635..173S}
{Sbarufatti}, B., {Treves}, A., \& {Falomo}, R. 2005, \apj, 635, 173

\bibitem[{{Scargle}(1982)}]{1982ApJ...263..835S}
{Scargle}, J.~D. 1982, \apj, 263, 835

\bibitem[{{Schramm} {et~al.}(1994){Schramm}, {Borgeest}, {Kuehl}, {von Linde},
  {Linnert}, \& {Schramm}}]{1994AAS..103..349S}
{Schramm}, K.-J., {Borgeest}, U., {Kuehl}, D., {von Linde}, J., {Linnert},
  M.~D., \& {Schramm}, T. 1994, \aaps, 106, 349

\bibitem[{{Shaw} {et~al.}(2012){Shaw}, {Romani}, {Cotter}, {Healey},
  {Michelson}, {Readhead}, {Richards}, {Max-Moerbeck}, {King}, \&
  {Potter}}]{2012ApJ...748...49S}
{Shaw}, M.~S., {et~al.} 2012, \apj, 748, 49

\bibitem[{{Shaw} {et~al.}(2013){Shaw}, {Romani}, {Cotter}, {Healey},
  {Michelson}, {Readhead}, {Richards}, {Max-Moerbeck}, {King}, \&
  {Potter}}]{2013ApJ...764..135S}
---. 2013, \apj, 764, 135

\bibitem[{{Shepherd}(1997)}]{difmap}
{Shepherd}, M.~C. 1997, in Astronomical Society of the Pacific Conference
  Series, Vol. 125, Astronomical Data Analysis Software and Systems VI, ed.
  G.~{Hunt} \& H.~E. {Payne} (San Francisco: ASP), 77

\bibitem[{{Sikora} {et~al.}(2005){Sikora}, {Begelman}, {Madejski}, \&
  {Lasota}}]{2005ApJ...625...72S}
{Sikora}, M., {Begelman}, M.~C., {Madejski}, G.~M., \& {Lasota}, J.-P. 2005,
  \apj, 625, 72

\bibitem[{{Sowards-Emmerd} {et~al.}(2005){Sowards-Emmerd}, {Romani},
  {Michelson}, {Healey}, \& {Nolan}}]{2005ApJ...626...95S}
{Sowards-Emmerd}, D., {Romani}, R.~W., {Michelson}, P.~F., {Healey}, S.~E., \&
  {Nolan}, P.~L. 2005, \apj, 626, 95

\bibitem[{{Stirling} {et~al.}(2003){Stirling}, {Cawthorne}, {Stevens},
  {Jorstad}, {Marscher}, {Lister}, {G{\'o}mez}, {Smith}, {Agudo}, {Gabuzda},
  {Robson}, \& {Gear}}]{2003MNRAS.341..405S}
{Stirling}, A.~M., {et~al.} 2003, \mnras, 341, 405

\bibitem[{{Thompson} {et~al.}(1992){Thompson}, {Djorgovski}, {Vigotti}, \&
  {Grueff}}]{1992ApJS...81....1T}
{Thompson}, D.~J., {Djorgovski}, S., {Vigotti}, M., \& {Grueff}, G. 1992,
  \apjs, 81, 1

\bibitem[{{Valtonen} \& {Wiik}(2012)}]{2012MNRAS.421.1861V}
{Valtonen}, M.~J., \& {Wiik}, K. 2012, \mnras, 421, 1861

\bibitem[{{Vermeulen}(1995)}]{1995PNAS...9211385V}
{Vermeulen}, R.~C. 1995, Proceedings of the National Academy of Science, 92,
  11385

\bibitem[{{Vermeulen} \& {Cohen}(1994)}]{1994ApJ...430..467V}
{Vermeulen}, R.~C., \& {Cohen}, M.~H. 1994, \apj, 430, 467

\bibitem[{{Vermeulen} {et~al.}(2003){Vermeulen}, {Ros}, {Kellermann}, {Cohen},
  {Zensus}, \& {van Langevelde}}]{2003AA...401..113V}
{Vermeulen}, R.~C., {Ros}, E., {Kellermann}, K.~I., {Cohen}, M.~H., {Zensus},
  J.~A., \& {van Langevelde}, H.~J. 2003, \aap, 401, 113

\bibitem[{{V{\'e}ron-Cetty} \& {V{\'e}ron}(2006)}]{VV12}
{V{\'e}ron-Cetty}, M.-P., \& {V{\'e}ron}, P. 2006, \aap, 455, 773

\bibitem[{{Vlahakis} \& {K{\"o}nigl}(2004)}]{2004ApJ...605..656V}
{Vlahakis}, N., \& {K{\"o}nigl}, A. 2004, \apj, 605, 656

\bibitem[{{Wagner} {et~al.}(1996){Wagner}, {Witzel}, {Heidt}, {Krichbaum},
  {Qian}, {Quirrenbach}, {Wegner}, {Aller}, {Aller}, {Anton}, {Appenzeller},
  {Eckart}, {Kraus}, {Naundorf}, {Kneer}, {Steffen}, \&
  {Zensus}}]{1996AJ....111.2187W}
{Wagner}, S.~J., {et~al.} 1996, \aj, 111, 2187

\bibitem[{{Wills} \& {Wills}(1976)}]{1976ApJS...31..143W}
{Wills}, D., \& {Wills}, B.~J. 1976, \apjs, 31, 143

\bibitem[{{Wright} {et~al.}(1983){Wright}, {Ables}, \&
  {Allen}}]{1983MNRAS.205..793W}
{Wright}, A.~E., {Ables}, J.~G., \& {Allen}, D.~A. 1983, \mnras, 205, 793

\bibitem[{{Zavala} \& {Taylor}(2005)}]{2005ApJ...626L..73Z}
{Zavala}, R.~T., \& {Taylor}, G.~B. 2005, \apjl, 626, L73

\end{thebibliography}
